\documentclass[twocolumn]{aastex631}
\usepackage{amsmath,amssymb}
\usepackage{tablefootnote}
\usepackage{longtable}
\shorttitle{Destruction of GMCs}
\shortauthors{D. Hollenbach et al}

\begin{document}

\newcommand{\beq}	{\begin{equation}}
\newcommand{\eeq}	{\end{equation}}
\newcommand{\beqa}{\begin{eqnarray}}
\newcommand{\eeqa}{\end{eqnarray}}
\newcommand{\beqs}	{\begin{displaymath}}
\newcommand{\eeqs}	{\end{displaymath}}
\newcommand{\beqas}	{\begin{eqnarray*}}
\newcommand{\eeqas}	{\end{eqnarray*}}
\newcommand{\shreq} {{\hspace*{-0.3cm}=\hspace*{-0.3cm}}}
\newcommand{\avg}[1]  {{\langle #1 \rangle}} 
\newcommand{\dis}{\displaystyle}
\newcommand\bit{\begin{itemize}}
\newcommand\eit{\end{itemize}}
\newcommand{\e}	{$^{-1}$}
\newcommand{\ee}	{$^{-2}$}
\newcommand{\eee}	{$^{-3}$}

\newcommand\calm{{{\cal M}}}
\newcommand\caln{{{\cal N}}}
\newcommand\calo{{{\cal O}}}
\newcommand\calr{{{\cal R}}}

\newcommand{\pbyp}[1]	{{\frac{}{}{\partial\hfil}{\partial#1}}}
\newcommand{\ppbyp}[2]	{{\frac{\partial#1}{\partial#2}}}

\newcommand{\ac}         {\alpha_{\rm cl}}
\newcommand{\alfven}    	{{Alfv$\acute{\rm e}$n}}
\newcommand{\alfvenic}  	{{Alfv$\acute{\rm e}$nic}}
\newcommand{\avir}      	{\alpha_{\rm vir}}
\newcommand\cs		{c_{\rm s}}
\newcommand\eff		{{\rm eff}}
\newcommand{\emax}       {\epsilon _{\rm max}}
\newcommand{\epff}       {\epsilon _{\rm ff}}
\newcommand{\eq}        {{\rm eq}}
\newcommand{\fmax}       {f _{M,\rm max}}
\newcommand{\kb}		{k_{\rm B}}
\newcommand\ma	    	{\calm_{\rm A}}
\newcommand{\Mc}       {M_{\rm cl}}
\newcommand{\Mch}       {M_{\rm cl,0.5}}
\newcommand{\Mcmax}       {M_{*,\rm max}}
\newcommand{\Mcmed}       {M_{*,\rm med}}
\newcommand{\Mcpm}       {M_{*,\rm pmm}}
\newcommand{\Mcu}       {M_{*,u}}
\newcommand{\Mcl}       {M_{*,\ell}}
\newcommand{\Mct}       {M_{*,{\rm T},av}}
\newcommand{\Ml}       {M_{c,\ell}}
\newcommand{\msun}   {M$_\odot$}
\newcommand{\msunm}     {\mbox{M}_\odot}
\newcommand\mug		{{$\mu$G}}
\newcommand{\muh}	{\mu_{\rm H}}
\newcommand{\Nh}		{N_{\rm H}}
\newcommand{\Nco}		{N_{\rm CO}}
\newcommand{\Aco}		{A_{\rm V,CO}}
\newcommand{\nh}		{n_{\rm H}}
\newcommand{\nhb}       {\bar n_{0}}
\newcommand{\nhei}		{n_{{\rm H,\,eq},i}}
\newcommand{\nhm}		{n_{\rm H,\,max}}
\newcommand{\nhn}		{n_{{\rm H},\,\nu}}
\newcommand{\nho}		{n_{\rm 0}}
\newcommand{\nhot}      {n_{\rm 0,2}}
\newcommand\pc		    {{\rm pc}}
\newcommand{\Sc}       {S}
\newcommand{\Sej}       {S_{\rm ej}}
\newcommand{\Sesc}       {S_{\rm esc}}
\newcommand{\Sejz}       {S_{\rm ej0}}
\newcommand{\Sunb}       {S_{\rm unb}}
\newcommand{\Sunbfn}       {S_{\rm unb,49}}
\newcommand{\Sunbeqfn}       {S_{\rm eq,49}(\xirunb)}
\newcommand{\Sunbtfn}       {S_{\rm t,49}(\xirunb)}
\newcommand{\Semb}       {S_{\rm emb}}
\newcommand{\Sblifn}       {S_{\rm bli,49}}
\newcommand{\Sflash}       {S_{\rm flash}}
\newcommand{\Sbli}       {S_{\rm bli}}
\newcommand{\Sflashfn}       {S_{\rm flash,49}}
\newcommand{\Scom}       {S_{\rm com}}
\newcommand{\Sdes}       {S_{\rm dest}}
\newcommand{\Sbld}       {S_{\rm b\ell d}}
\newcommand{\Sbleq}       {S_{\rm b\ell stall}}
\newcommand{\Seqd}       {S_{\rm stalld}}
\newcommand{\Seqfn}      {S_{\rm stall,49}}
\newcommand{\SHII}       {S_{\rm flash,49}}
\newcommand{\Scra}     {S_{\rm cr1}}
\newcommand{\Scrb}       {S_{\rm cr2}}
\newcommand{\Scrc}       {S_{\rm cr3}}
\newcommand{\Scrd}       {S_{\rm cr4}}
\newcommand{\Scre}       {S_{\rm cr5}}
\newcommand{\Scrf}       {S_{\rm cr6}}
\newcommand{\Scrg}       {S_{\rm cr7}}
\newcommand{\Scrh}       {S_{\rm cr8}}
\newcommand{\rms}       	{{\rm rms}}
\newcommand{\tff}		{t_{\rm ff}}
\newcommand{\tffb}       {\bar t_{\rm ff}}
\newcommand{\tffo}		{t_{\rm ff,0}}
  
\newcommand{\va}		{v_{\rm A}}
\newcommand{\vao}		{v_{\rm A0}}

\newcommand{\blue}[1]{{\textcolor{blue}{#1}}}
\newcommand{\red}[1]{{\textcolor{red}{#1}}}
\newcommand{\mage}[1]{{\textcolor{magenta}{#1}}}
\newcommand{\cf}        {\mage{CF}\ }
\newcommand{\cfm}       {\mage{CFM}\ }

\newcommand{\cl}        {{\rm cl}}
\newcommand{\CO}        {{\rm CO}}
\newcommand{\cotz}      {{c,\,1-0}}
\newcommand{\cootz}     {{c0,\,1-0}}
\newcommand{\emsun}     {\msunm}
\newcommand{\epsff}     {\epsilon_{\rm ff}}
\newcommand{\epsbr}     {\epsilon_{\rm br}}
\newcommand{\epsgmc}     {\epsilon_{\rm GMC}}
\newcommand{\epsffmt}     {\epsilon_{\rm ff,-2}}
\newcommand{\epsmax}    {{\epsilon_{\rm max}}}
\newcommand{\epssfc}     {\epsilon_{\rm SFC}}
\newcommand{\esc}        {{\rm esc}}
\newcommand{\euv}       {{\rm EUV}}
\newcommand{\fmin}       {f_M}
\newcommand{\fto}       {f_{31}}
\newcommand{\fuv}       {{\rm FUV}}
\newcommand{\gmc}       {{\rm GMC}}
\newcommand{\gmcu}       {{\gmc,u}}
\newcommand{\ionz}       {{\rm ion}}
\newcommand{\ir}        {{\rm IR}} 
\newcommand{\mds}				{{\dot m}_*}
\newcommand{\mdu}				{{\dot m}_u}
\newcommand{\msh}               {{m_{*h}}}
\newcommand{\msu}               {{m_{*u}}}
\newcommand{\Mds}				{{\dot M}_*}
\newcommand{\Mdsb}				{\overline{\dot M}_*}
\newcommand{\Mdst}				{{\dot M}_{*T}}
\newcommand{\Mcs}				{M_{6}}
\newcommand{\mso}       {M_{s0}}
\newcommand{\mtc}       {{M_{T,\,\rm Col}}}
\newcommand{\ncu}       {{\caln_{cu}}}
\newcommand{\nclmax}       {{\caln_{\rm cl,\,max}}}
\newcommand{\ncuc}       {{\caln_{cu,\,\rm Col}}}
\newcommand{\nds}				{{\dot\caln_*}}
\newcommand{\ndsh}				{{\dot\caln_{*h}}}
\newcommand{\ndu}				{{\dot\caln_u}}
\newcommand{\nhas}      {\avg{\nh^{1/2}}_{M6}}
\newcommand{\nl}        {{n_{\rm L}}}
\newcommand{\otz}       {{1-0}}
\newcommand{\pif}       {P_{\rm IF}}
\newcommand{\rc}               {{r_{cs}}}
\newcommand{\rcs}               {{r_{cs}}}
\newcommand{\rcb}               {{r_{cb}}}
\newcommand{\rCO}      {r_{\rm CO}}
\newcommand{\rcr}       {r_{\rm cr}}
\newcommand{\rcz}               {{r_{c0}}}
\newcommand{\rcunb}               {{r_{\rm unb}}}
\newcommand{\rcom}      {{r_{\rm com}}}
\newcommand{\rcej}      {{r_{\rm ej}}}
\newcommand{\rceq}        {{r_{\rm stall}}}
\newcommand{\rs}        {{r_s}}
\newcommand{\rsf}       {r_{s,f}}
\newcommand{\Rc}          {{R_c}}
\newcommand{\sfc}       {{\rm SFC}}
\newcommand{\sfn}       {{S_{49}}}
\newcommand{\sofn}      {S_{0,49}}
\newcommand{\sfnh}      {{\avg{s_{49}}_h}}
\newcommand{\tco}       {{$^{12}$CO}} 
\newcommand{\teq}       {t_{\rm stall}}
\newcommand{\tcom}     {t_{\rm com}}
\newcommand{\tcomf}     {t_{{\rm com},f}}
\newcommand{\tcz}       {{t_{c0}}} 
\newcommand{\tffh}		{\avg{\tff^{-1}}^{-1}}
\newcommand{\tffu}		{t_{{\rm ff},u}}
\newcommand{\thej}       {{\theta_{\rm ej}}}
\newcommand{\thecs}       {{\theta_{cs}}}
\newcommand{\Thecs}     {{\Theta_{cs}}}
\newcommand{\thesc}       {{\theta_{\rm esc}}}
\newcommand{\thecf}       {{\theta_{cs,f}}}
\newcommand{\thescf}       {{\theta_{{\rm esc},f}}}
\newcommand{\theq}       {{\theta_{\rm stall}}}
\newcommand{\thunb}     {{\theta_{\rm unb}}}
\newcommand{\thcom}     {{\theta_{\rm com}}}
\newcommand{\tms}       {t_{\rm ms}}
\newcommand{\ttt}       {{(3-2)}}
\newcommand{\vesc}      {v_{\rm esc}}
\newcommand{\xir}       {{\xi}}
\newcommand{\xirs}       {{\xi_s}}
\newcommand{\xirS}       {{\xi_S}}
\newcommand{\xirc}       {{\xi_{cs}}}
\newcommand{\xircs}       {{\xi_{cs}}}
\newcommand{\xircf}       {{\xi_{cs,f}}}
\newcommand{\xircz}       {{\xi_{c0}}}
\newcommand{\xirunb}       {{\xi_{\rm unb}}}
\newcommand{\xircom}       {{\xi_{\rm com}}}
\newcommand{\xirej}       {{\xi_{\rm ej}}}
\newcommand{\xireq}       {{\xi_{\rm stall}}}
\newcommand{\xires}     {{\xi_{\esc}}}
\newcommand{\xiresf}     {{\xi_{\esc,f}}}
\newcommand{\xirsf}      {\xi_{s,f}}
\newcommand{\xirsto}     {\xi_{\rm St,0}}

\font\cmbi=cmmib10 
\font\tenbi=cmmib8
\newfam\bifam \def\bi{\fam\bifam\tenbi} \textfont\bifam=\tenbi
 \newcommand{\drs}		{{\Delta r_s}}
\newcommand{\unb}		{{\rm unb}}
\def\vecg               {{\bi{g}}}
\def\vecr               {{\bi{r}}}
\def\vecR               {{\bi{R}}}
\def\vecu               {{\bi{u}}}
\def\vecz               {{\bi{z}}}
\def\vecZ               {{\bi{Z}}}
\def\vecvarpi            {{\bi{\varpi}}}
\newcommand{\ab}		{\alpha_{\rm B}}
\newcommand{\bli}       {{\rm bli}}
\newcommand{\capp}	    {{\rm cap}}
\newcommand{\cii}		{c_{\rm II}}

\newcommand{\cm}        {{\rm cm}}
\newcommand{\degree}	{$^\circ$}
\newcommand{\ej}        {{\rm ej}}
\newcommand{\emb}		{{\rm emb}}
\newcommand{\eamo}       {\epsilon_{a,\max,-1}}
\newcommand{\emt}       {\epsilon_{-2}}  
\newcommand{\evap}      {{\rm evap}}
\newcommand{\fii}		{F_{\rm II}}
\newcommand{\fiio}		{F_{\rm II,0}}
\newcommand{\fion}		{f_{\rm ion}}
\newcommand{\flash}		{{\rm flash}}
\newcommand{\gas}       {{\rm gas}}
\newcommand{\hii}       {{\rm ion}}
\newcommand{\ini}        {{\rm emb}}
\newcommand{\init}      {{\rm init}}
\newcommand{\ism}		{{\rm ISM}}
\newcommand{\loss}      {{\rm loss}}
\newcommand{\myr}       {{\rm Myr}}
\newcommand{\mii}		{M_{\rm ion}}
\newcommand{\miiom}		{M_{\rm ion,\,\Omega}}
\newcommand{\mevom}      {M_{\rm evap,\Omega}}
\newcommand{\mionom}      {M_{\rm ion,\Omega}}
\newcommand{\msom}      {M_{s,\Omega}}
\newcommand{\nii}		{n_{\rm II}}
\newcommand{\nsn}       {\caln_{\rm SN}}
\newcommand{\rhoii}		{\rho_{\rm II}}
\newcommand{\niio}		{n_{\rm II,0}}
\newcommand{\pii}		{P_{\rm II}}
\newcommand{\phiP}	    {\phi_{\rm P}}
\newcommand{\phiii}	    {\phi_{\rm II}}
\newcommand{\phieff}	{\phi_{\rm P,eff}}
\newcommand{\phiiieff}	{\phi_{\rm II,eff}}
\newcommand{\pkm}       {\psi_{\rm KM}}
\newcommand{\rad}       {{\rm rad}}
\newcommand{\rhoiio}	{\rho_{\rm II,0}}
\newcommand{\rch}		{r_{\rm ch}}
\newcommand{\rchpc}	    {r_{\rm ch,\,pc}}
\newcommand{\ropc}	    {r_{0,\rm pc}}
\newcommand{\rii}		{r_{\rm II}}
\newcommand{\rst}		{r_{\rm St}}
\newcommand{\rsto}		{r_{\rm St,0}}
\newcommand{\sn}        {{\rm SN}}
\newcommand{\term}      {{\rm term}}
\newcommand{\vii}		{v_{\rm II}}
\newcommand{\viieff}	{ v_{\rm II,eff}}

\newcommand{\minl}      {M_\init(<\thecs)}
\newcommand{\miong}      {M_\ionz(>\thecs)}
\newcommand{\sto}		{{\rm St,0}}
\newcommand{\viieq}	  	{v_{\rm II,\,stall}}

\newcommand{\cav}			{{\rm cav}}
\newcommand{\cen}       {{\rm cen}}
\newcommand{\cone}		{{\ionz,\,\rm cone}}
\newcommand{\murcs}		{{\mu_{r,cs}}}
\newcommand{\murcsd}		{\dot\mu_{r,cs}}
\newcommand{\nedge}     {n_{\rm edge}}
\newcommand{\rhoiics}		{\rho_{{\rm II},cs}}
\newcommand{\rhoiit}		{\rhoii(\theta)}
\newcommand{\rhoiitcs}		{\rhoii(\thecs)}
\newcommand{\rsth}         {{r_s(\theta)}}
\newcommand{\sdmto}     {\sigma_{d,-21}}
\newcommand{\tion}      {t_{\rm ion}}
\newcommand{\vecn}      {{\bi{n}}}
\newcommand{\zcs}       {z_{cs}}
\newcommand{\Zcs}       {Z_{cs}}




\title{GMCs and Star Formation in the Galaxy: I. Mass Loss from a GMC Induced by an HII Region}


\author{David Hollenbach}
\address{15 Cerro Blanco Rd, Lamy, New Mexico, USA}
\email{davidjhollenbach@gmail.com}
\author{Antonio Parravano}
\address{Universidad de Los Andes, Centro De Física Fundamental, Mérida 5101a, Venezuela}
\author{Christopher F. McKee}
\affil{Departments of Physics and of Astronomy, University of California, Berkeley, CA 94709, USA}

\date{Accepted by ApJ 10/24/2025. Received 09/06/2025}



\label{firstpage}

\begin{abstract}
The destruction of Giant Molecular Clouds  is a key component in galaxy evolution. We  theoretically model 
 the destruction of GMCs by HII regions, which evaporate ionized gas and eject neutral gas during their expansion. HII regions follow one of three tracks, depending on the EUV luminosity, $S$, of the ionizing OB association: the expansion can stall
inside the cloud; it can break out, forming a blister (champagne)  flow; or, for $S>S_{\rm com}$, it can result in the formation of a cometary cloud.   We present results for the accumulated mass loss,
$M_{\rm loss}(t)$, and the final
mass loss, $M_{{\rm loss},f}$, by evaporation and ejection for a range of cloud masses ($10^4<M<10^{7}$ M$_\odot$), cloud surface densities
($50<\Sigma<1000$ M$_\odot$ pc$^{-2}$), OB
association luminosities ($10^{44}<S<10^{52}$ s$^{-1}$), and off-center position of the association. 
We do not consider starbursts; our neglect of radiation pressure restricts our treatment to $S<10^{52} [(M/10^6\,\mbox{\msun})^{0.3}/(\Sigma/ 100\,\mbox{\msun\ pc\ee})]$ s$^{-1}$, and our neglect of gravity restricts $(M/10^6\,\mbox{\msun})(\Sigma/ 100\,\mbox{\msun\ pc\ee}) \la 10$.
We find that $M_{{\rm loss},f}$ for the range $0.1 \la M_{{\rm loss},f}/M \la 0.7$ , is proportional to $S^p$, 
where $p\sim 0.45-0.75$ depends on $M$, $\Sigma$,
and association position. We find analytic fits to $S_{\rm com}$ as a function of $\Sigma$, $M$, and association position.  $S>\Scom$ associations destroy at least 70\% of the initial cloud. We find a
critical cloud mass $M_{\rm survive}$ above which clouds never become cometary and lose $\la$ 70\% of their mass via a single association. 
Low mass clouds mostly lose mass via ejection of neutral gas.


\end{abstract}



\section{Introduction}

Galaxies are factories  converting gas and dust to stars and planets.  Giant Molecular Clouds (GMCs) play the dominant role in this production, but they are not efficient players in the assembly.   During their lifetimes they typically convert less than about 10 percent of their initial mass to stars (e.g., \citealp{chev20}). A key reason for this low efficiency is that the massive stars produced by GMCs cause them to self-destruct in short order.  Associations of massive stars form in dense cores in GMCs, rapidly break out of the remaining gas in the core, and expand into the ambient GMC. Numerous authors, beginning with \citet{blit80}, have focused on HII regions as the dominant destruction mechanism. \citet{lope14} showed that observations of HII regions in the Magellanic Clouds support this conclusion: In almost all cases the pressure of the warm, photoionized gas dominated the pressures of the hot gas produced by stellar winds and supernovae and the pressure due to radiation.

In this paper we analytically and numerically examine in detail the mass loss from a GMC due to the EUV luminosity (i.e., the ionizing luminosity) from an OB association that forms  in the cloud.\footnote{We use the terms ``cluster" and ``association" interchangeably. Clusters are sometimes defined as gravitationally bound collections of stars, whereas associations are generally taken to be unbound. However, as discussed by \citet{krum19}, it is generally not possible to determine if young clusters/associations are bound. We adopt the term ``association" to avoid possible confusion in the notation for labels like ``cl" referring to clusters on the one hand and clouds on the other.}   For application to Milky Way GMCs we rely on the results of 
Rosolowsky et al (in preparation), who used high-resolution CO data 
to establish the relation between the cloud radius, $R_c$, and the cloud mass, $M$, in the Galaxy. 
Here we follow the EUV-induced photoevaporation and shell ejection from a cloud of mass $M$ as a function of the ionizing photon luminosity, $S$, of a single association, of the placement of that association in the cloud, and of the surface density $\Sigma=M/\pi R_c^2$ of the cloud.  In Parravano et al. (in preparation) we use our results to
determine the lifetimes of GMCs once massive star formation commences.   

Various authors have made previous  contributions to the study of the effect of the EUV luminosity of a single association on 
mass loss from a GMC. We first discuss analytical and  semi-analytical models like the ones we present in this paper.  \citet{spit78} set down the basic equations for the evolution of an embedded HII region as it expands into its natal cloud.  However, being embedded, the growing HII region induced no mass loss from the cloud, only a transfer of mass from neutral state to confined ionized gas.  \citet{whit79} provided an analytic solution for ``champagne", or blister,  evaporative flow from a GMC.  An O star that forms near the surface of a GMC has its HII region break out the near the surface, and subsequently the hot ionized HII region can expand outwards in that direction, exiting the cloud to the ISM at
speeds  of order the sound speed in the ionized gas, $\cii\simeq 10$~km~s\e,  but accelerating via pressure gradients to speeds $\gg\cii$.
\citet{will97} (hereafter WM97), used Whitworth's  results but stopped 
evaporation either when the massive stars in the association supernovaed after about 4 Myr, or
(for luminous associations) after the shell had 
traversed a distance $\sim R_c$, which they termed the ``disruption" of the cloud.  \citet{matz02} generalized to a hemispherical shell
expanding into the cloud, pointing out that the length scale $L$ in Whitworth (who assumed a cubic shape) was the hemisphere diameter; he also included the inertia of the swept-up shell in the dynamical equations.
\citet{krum06} included the ejection of the neutral shells that were
driven to speeds greater than escape speed.
Although they assumed hemispherical expansion,
and therefore photoevaporative escape of the HII plasma during the shell evolution, they essentially placed the association at cloud center. Instead of constant density cloud, they assumed the density fell as $R^{-1}$,
where $R$ is distance to cloud center. If the shell was moving faster than the escape speed when the shell reached the cloud surface at $R_c$, they added that mass loss to the evaporated mass loss as ``ejected" neutral mass.   They pointed out this was an important source of mass loss and the dominant term in the cloud lifetime for low mass ($<10^5$ M$_\odot$) clouds.

In parallel with these analytical and semi-analytical developments, numerical hydrodynamical models were applied to the evolution of the expanding HII region around either a single O star or an association.   
\citet{ten79} produced the first numerical model of champagne flow, and further developed it in \citet{bod79},  and \citet{ten82}.  In addition, \citet{york82} and \citet{york89}  constructed hydrodynamical models that placed the O star at various depths in a GMC, and also examined the effect of gravitational collapse of the cloud on the champagne flow.     
\citet{walc12} used a 3D SPH code to follow the evolution of the HII region produced by a single O7 star placed at the center of a fractal $10^4$ M$_\odot$ cloud.   Interestingly, they found that the evolution of the mean ionization front followed the simple analytic treatments that used the mean density, and that the outflow is not very dependent on the fractal dimension of the clumpy cloud.   Like \citet{krum06} above, they found that the expulsion or ejection of the neutral shell dominated the dispersal time for this cloud, and the cloud was dispersed by this process in 1-2 Myr after the turn-on of the star.
Although this Introduction does not discuss the vast literature of simulations of GMCs that include cloud formation and the effects of the formation of multiple associations on the cloud, we mention two sets of such studies here, \citet{dale12,dale13,dale14}
 and \citet{kim18}, because they too find that low mass clouds have significant mass loss from the ejection of neutral shells, whereas high-mass clouds are mostly destroyed by photoevaporation of ionized gas.

In this paper we address the problem of the evolution of an HII region powered by a stellar association embedded in a GMC.  In Section 2 we describe and justify our model in detail, but here we simply note that we approximate the evolution with a spherical, constant density cloud where the association is placed off-center. We include a (magnetic or turbulent) pressure gradient to support the cloud.   We ignore radiation pressure, stellar winds and supernovae.  We treat both the evaporation of ionized gas and the ejection of neutral gas as mass loss mechanisms.

Our model is clearly simplified, but it enables us to achieve semi-analytic and analytic results for the evolution of an HII region over a wide variety of model parameters. Our solutions depend on five parameters: the cloud mass, $M$, the cloud surface density, $\Sigma$, the ionizing luminosity of the association, $S$, the ambient interstellar pressure, $P_{\rm ISM}$, and the location of the association in the cloud. 
On average, the surface density of GMCs in the Milky Way is approximately constant 
(\citealp{lars81, roma10}, Rosolowsky et al in preparation), so only four parameters are needed to describe the evolution of HII regions in Galactic clouds.

In Section 3 we summarize the properties of GMCs and the evolution of stellar associations.
Section 4 gives the results we obtain from numerial integration of the equations describing our model, whereas Section 5 contains analytic approximations to these results.
In Section 6 we
focus on Milky Way clouds that follow the observed  mass-size  relation, and in Section 7 we generalize the
results to GMCs of a large range of surface densities, especially relevant to external galaxies and unusual clouds in the Milky Way.  Section 8 gives a simple parametric fit to the final mass loss as a function of $\Sigma$, $M$, and association position.
 Appendix A discusses the fraction of EUV photons absorbed by gas and not dust.  Appendix B presents the condition that radiation pressure can be neglected.
 Appendix C gives the ISM pressure at a typical point in the Milky Way. Appendix D discusses the photoevaporation rate, and Appendix E provides  a prescription for  an approximate analytic solution to cloud mass loss as a function of the input parameters.

\section{Overall Description of the Model}
\label{sec:overall}

\begin{figure*}

\hbox{\includegraphics[scale=1.0,angle=0]{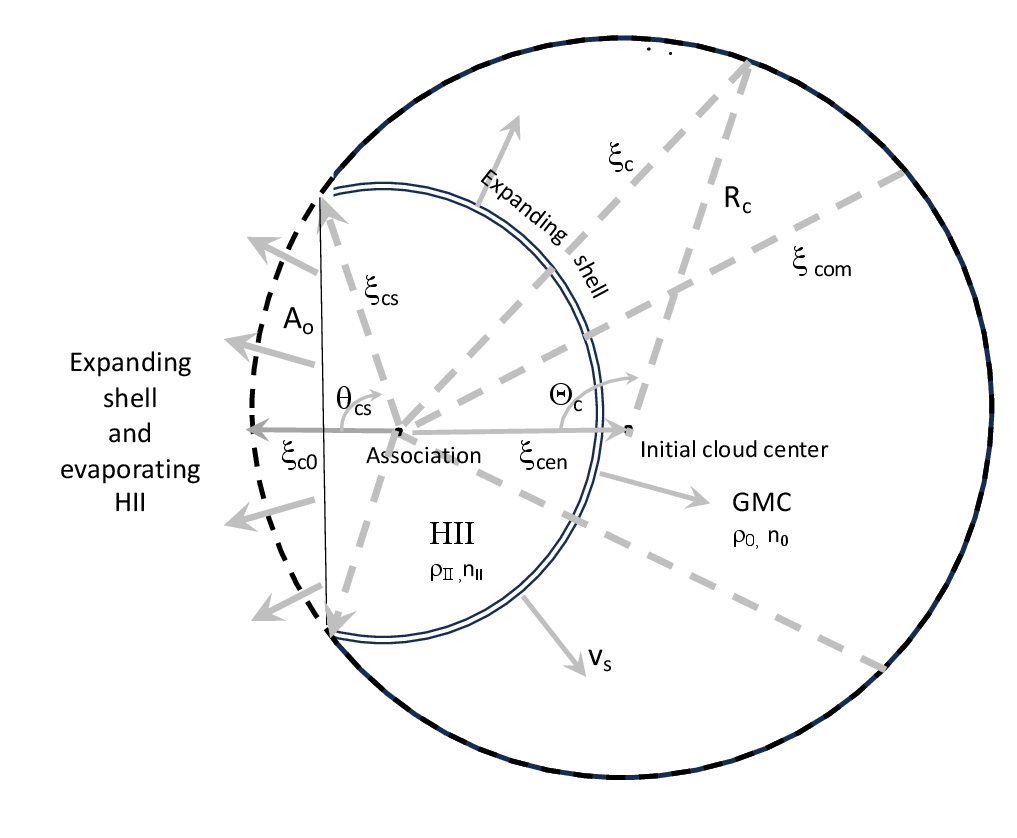}}

\caption{The GMC is assumed spherical with radius $\Rc$; the normalized distance from the association to the cloud surface is $\xi_c=r_c/R_c$.  The number density of H nuclei in the cloud (HII region) is $\nho$ ($\nii$) and the mass density is $\rho_0$ ($\rhoii$). In our numerical work, $\nii$ and $\rhoii$ are functions of $\theta$,  but in our analytic work we take them to be independent of $\theta$. Similarly, the shell is spherical (as pictured above)  in our analytic work, but non-spherical in our numerical work.
 The association is located at a normalized distance $\xircz=r_{c0}/ R_c$ from the nearest cloud surface and at a normalized distance $\xi_\cen=r_\cen/R_c=R_a/R_c$ from the cloud center.  Ionizing radiation from the association creates an HII region that drives a shell of neutral gas into the cloud with a normalized radius $\xi_s=r_s/R_c$ and velocity $v_s$. 
 Once the HII region transitions from fully embedded to a blister (pictured), the shell intersects the cloud surface at a
dimensionless distance $\xirc\equiv \rc/\Rc$ from the association, and the angle this ray makes with the line going through the association and cloud center is $\theta_{cs}$. $A_o$ is the area of the opening to the ISM.
The partially enclosed HII region lies between $A_o$ and the internal shell surface (area $A_s$), and has a mass $M_{\rm ion,os}$.
 Gas is ejected and evaporated from the cloud out of the opening $A_o$. 
We assume  the cloud transitions to cometary cloud when $\thecs=150^\circ$.
}
\label{fig-3} 
\end{figure*}

Figure \ref{fig-3} shows the basic setup of our cloud model. 
Distances from the initial cloud center are denoted by $R$; the association is located at $\vecR_a$ and the initial cloud radius is $R_c$. Distances from the association are denoted by $r$, which we express in dimensionless form as
\begin{equation}
\xir\equiv \frac{r}{ \Rc}.
\label{eq:xi}
\end{equation}
The cloud surface is located at a distance $r_c(\theta)$ from the association (the subscript ``c" denotes a position on cloud surface), where $\theta$ is the angle between $\vecr$ and $\vecR_a$.
The minimum distance from the association to the cloud surface is at $\theta=0$, and we define that as $r_{c0}=r_c(\theta=0)$.
In dimensionless form, this is
\begin{equation}
    \xircz\equiv \frac{\rcz}{\Rc} =1-\frac{R_a}{R_c}\equiv 1-\xi_\cen,
    \label{eq:xircz}
\end{equation}
where the subscript ``cen" refers to the normalized distance to the cloud center.
The shell intersects the surface at a radius $\rcs$ (the subscript ``cs" refers to the intersection of the shell with the surface of the cloud) and an angle $\theta_{cs}$.
The normalized radius $\xircs$ is related to $\theta_{cs}$ by
\beq
\cos\theta_{cs}\equiv \murcs=\frac{2\xircz-\xircz^2-\xi_{cs}^2}{2\xi_{cs}(1-\xircz)}.
\label{eq:cos}
\eeq
The circular opening to the ISM has an area $A_o=\pi (\rcs \sin\theta_{cs})^2$.

We follow the evolution of an HII region
produced by an association that is off center in the cloud, located at a distance
0.2-0.4 $R_c$ from the cloud surface--i.e., corresponding to $\xircz=0.2-0.4$.  
The HII region drives a shell of neutral gas into the cloud and, in some directions, away from the cloud (see Fig. \ref{fig-3}).
We consider associations with a wide range of ionizing photon luminosities, $S$ (see Section 3.2), initially embedded in GMCs of a wide range of cloud masses, 
$ 10^4 \la M\la10^7$ M$_\odot$ and a moderate range of cloud surface densities, $\Sigma=50-1000$~\msun\ pc\ee.

\subsection{Approximations}

Our model includes  nine  key  approximations:

1. We assume that the cloud is initially spherical with a density that is independent of radius in the cloud. 
In reality clouds are clumpy and may have an overall density gradient,  but our results should still approximately apply \citep{krum06,walc12}.
The parameter
$\xircz$ can be adjusted to roughly account for these effects.  For example, clumpy clouds have low density channels to the
cloud surface, so that embedded HII regions can more easily break out to the surface.  Similarly, filamentary or non-spherical clouds lead to star formation occurring closer to a ``surface".  Reducing $\xircz$ mimics these effects on evaporated or ejected mass loss.  
In Section 4.5 we compare the results of our model to the high resolution 3D hydrodynamical calculations of various clumpy clouds \citep{walc12} and find satisfactory agreement.
As for the assumption that the average density is independent of radius, we note that had we assumed an $R^{-1}$ density gradient, for example, the value of $\xircz$ would have increased  for a fixed enclosed mass,  but the column density to the surface would not have changed much, thereby countering the effect of the change in location of the association.

2.  We assume that the structure of the cloud outside the HII region is constant in time; in particular, the cloud is initially in hydrostatic equilibrium.  The cloud pressure increases with decreasing $R$. We follow the evolution of the association for the effective ionization lifetime of the association (Section \ref{sec:evosc}). 
Neglecting the time variation of the cloud structure should be a good approximation for associations large enough to fully sample the IMF (those with masses $M_a\ga 2000$~\msun--see Eq. \ref{eq:tion} below), since they have effective ionization lifetimes $\sim 4$~Myr. 
Our treatment becomes increasingly approximate for associations with lower masses and longer lifetimes.

3. We neglect radiation pressure, supernovae and stellar winds since they are not the dominant destruction mechanisms for GMCs similar to those in the Galaxy. 
The observational evidence for this statement is given by \citet{lope11} and \citet{lope14} for clouds in the LMC, which are not that different from those in the Galaxy.  Radiation pressure is discussed in Appendix \ref{app:rp}.  There we show that radiation pressure is significant only for very large associations in massive clouds with surface densities generally exceeding those in Galactic GMCs but typical of starbursts; we therefore do not consider starbursts here. \citet{chev22} show that supernovae do not dominate GMC destruction. For stellar winds, \citet{mcke84} argued that mass input from photoevaporating clumps in stellar wind bubbles would lead to cooling and reduce the size of the bubbles. \citet{matz02} pointed out that in blister HII regions hot gas escapes the cloud.  Simulations by \citet{mack15} and \citet{lanc21} show that most of the wind energy is radiated away due to mixing of the shocked wind gas with the surrounding HII region.                     We conclude that stellar winds can also be neglected in estimating the lifetimes of GMCs. Further discussion of the relative unimportance of radiation pressure, stellar winds and supernovae in GMC destruction is given by \citet{chev22}.

4. We assume that the association is born suddenly, with the full value for its ionizing luminosity. In fact, associations form over a time exceeding the free-fall time, $\tff$. Analysis of the data of \citet{koun18} on the Orion Nebula Cluster shows that about half the stars form in $3\tff$, or about 2 Myr \citep{krum20}. The data of \citet{pris19} show that over half the stars in NGC 6530 formed in less than 1 Myr. A formation time of 1-2 Myr is a significant fraction of the 4 Myr lifetime of a massive starburst. In both cases these are upper limits, since the significant uncertainties in the ages contribute to the measured dispersion. Furthermore, these star formation times are based on observations of low-mass stars, whereas the massive stars could form in a shorter time interval since they form in very dense gas (e.g.,\citealp{plum97}). Correspondingly, we assume that the association dies suddenly. We note that this approximation does not affect small associations, which are dominated by a single massive star.

5. We place the association  in the cloud interior
 at $\xircz=0.2$, 0.3 or 0.4 times the cloud radius from the surface of the initial GMC. This differs from most (non-numerical) treatments of the evolution of HII regions, which place the HII region at the center or edge of the cloud.    
Note that at $\xircz=0.2$, the association is at the half mass radius in the assumed constant density cloud, and at $\xircz=0.4$ only 22\% of the cloud mass lies inside the association.  One might be
tempted to therefore place our associations at the half mass point, $\xircz=0.2$. In fact, star formation is suppressed in the outer layers of GMCs \citep{mcke07}, so the typical association would form at a somewhat greater distance from the surface, $\xircz\sim 0.3-0.4$.  
Our parameter $\xircz$ should be viewed not only
as an average position for an association (since associations orbit in the cloud), but also as taking into account the inhomogeneous and non-spherical nature of the cloud.

6. We assume  that the
association is stationary with respect to the GMC.   
This is a good approximation initially, since the association is born at the same velocity as the
gas from which it forms. However, the relative velocity between the association and the cloud
builds up in time because the cloud is turbulent and the gas feels pressure forces whereas the association feels only the gravity of the GMC. As a result,
associations---especially long-lived associations in low mass clouds---orbit substantially in the cloud, which can affect the resulting cloud mass loss. 
\citet{matz02} suggested that motion of the association would be important when the free-fall time was shorter than the lifetime of the association, $\tion$. For the parameters he adopted, this restricted the validity of his analysis to $M\ga 4\times 10^5\;\msunm$. The density we have adopted (Eq. \ref{eq:nh} below) is about half of his value, so this condition is somewhat less restrictive in our case, $M\ga 1.0\times 10^5\;\msunm$. 
\citet{mack15} have simulated the motion of a massive star through a uniform cloud. Their results imply
that the ionization front in the direction of motion is given by the Str\"omgren radius evaluated at a
density $\rhoii$ given by pressure balance, $\rhoii\cii^2 = \rho_0 v_*^2$, provided the stellar velocity
$v_*$ is supersonic relative to the ambient medium but less than $2\cii$ so that a D-type ionization front (IF) can form.
The radius of the IF grows as the angle relative to the direction of motion increases. Remarkably
enough, they find that the total mass of the shell swept up by the IF is almost exactly the same
as that for a static star. This result shows that motion of the association has a greater effect
on the geometry of the swept up gas than on the amount, so the stationary approximation may
continue to give qualitatively correct results for cloud destruction even for moving associations in clouds with
masses less than $10^5$~\msun.

 7.  We  assume that the  neutral gas is swept up by the HII region into a thin shell moving radially outward from  the association.  In the absence of gravity this is a very good approximation.  We initially included the effect of the gravity of the GMC on the shell, but found this to be unimportant (in accord with \citealp{oliv21}) and so have omitted this effect. 
 However, we have included two effects of gravity, which are also discussed in the subsequent paragraph: First, gravity determines the pressure inside the cloud since we assume that the cloud is initially in hydrostatic equilibrium. Second,  it determines the escape speed from the cloud, which declines with time due to mass loss.  If the escape speed is much higher than the sound speed in the HII plasma, the photevaporative mass loss rate is significantly reduced and we omit this part of parameter space from our study.

8. We include an improved treatment of the dynamics of the expanding HII region and the neutral shell that surrounds the HII region.  First, the shell is driven not only by the thermal pressure of the HII region, but also by the rocket effect of gas evaporating off the inner shell surface--i.e., the gas just behind the ionization front, or IF \citep{matz02}.  The improvement we introduce is that rocket effect depends on the speed of the ionized outflow from the IF, which in turn depends on the shape of the HII cavity. Second, we include the effect of the pressure in the cloud.   The cloud pressure can cause the shell to stall in the cloud, especially in directions toward the cloud center where the pressure is higher.  Third, because our associations are off-center, partial ejection of shells occurs in directions away from the cloud center, even when the shell might stay embedded in the direction toward the cloud center.   If an association is luminous enough to drive a part of the shell out of the cloud,  we find that it quickly accelerates to escape speed  since the ambient interstellar pressure is much lower than that in the GMC, and as a result it is ejected from the cloud. We treat both evaporated  ionized  gas and ejected neutral shells as mass loss from the cloud.  We find that in many circumstances (but especially for low-mass clouds)  most of the mass lost from the GMC is ejected neutral gas that will be subject to photodissociation by the associations in the cloud and the interstellar radiation field.

9. We distinguish between ejected neutral shells that merge with the ISM 
(mass loss from cloud) and the remnant (cometary) cloud that
is pushed away from the assocation via the rocket effect of photoevaporation.
Associations with high EUV luminosities can drive shells to the opposite side of the cloud from the association.  As noted by WM97, in this case photoevaporation (with an assist from self-gravity) will drive the remaining cloud into the cometary shape treated by \citet{bert90}. Here we adopt $\theta=150^\circ$
 as the critical angle beyond which the ``ejected" neutral cloud material remains in a coherent molecular structure that we treat as the remnant of the cloud, not mass lost from the cloud. However, this cometary cloud is subject to photoevaporation, which is counted as mass lost from the cloud.

These nine approximations enable us to integrate the equation of motion along radial trajectories as a function of $\theta$. With the further assumption that the motion of the neutral shell is independent of $\theta$, it is possible to obtain approximate analytic solutions with a single free (but physically constrained) parameter that can be adjusted to obtain improved agreement with the numerical solutions. Our prescription for an analytic solution gives the
   mass loss as a function of the ionizing luminosity, $S$, the association position, $\xircz$, the cloud mass, $M$, the cloud surface density, $\Sigma$, and the time since the association was born, $t$.
  Analytic solutions are essential for 
  semi-analytic treatments of cloud destruction by all the HII regions that form in a cloud.  They also show the scalings of various dynamical quantities with these parameters.

\begin{longtable*}{ll}
\caption{Commonly Used Symbols}
\label{tab:comm}\\
\hline
Symbol & Definition\\
 \hline
 \endfirsthead
 \hline
 \multicolumn{2}{c}{Commonly Used Symbols--continued}\\
 \hline
 Symbol & Definition \\
 \hline
 \endhead

\hline
\endfoot

$A_o$&Area of opening to the ISM during blister stage (below Eq. \ref{eq:cos}; Fig. \ref{fig-3})\\
$A_s$&Area of shell inside cloud illuminated by association (Eq. \ref{eq:mdhii})\\
$\cii$&Isothermal sound speed of ionized gas $=11.1T_4^{1/2}$~km s\e; we take $T_4=1$\\
$\fion$& The fraction of ionizing photons absorbed by the gas (as opposed to dust) in an HII region (Eq. \ref{eq:fion})\\
 $M$&Initial cloud mass measured in \tco\ (1-0); $M_6=M/(10^6$~\msun)\\
 $M_a$&Initial mass of stars in association; $M_{a,\max}=\epsilon_{a,\max} M$, maximum mass of an association in a cloud of mass $M$ \\
 
 $M_{\rm ej}$&Mass of gas in the neutral shell that is ejected from the cloud by the pressure of the ionized gas ($\theta<\thecs$)  \\
 $M_\evap$&$M_\ionz -M_{\rm ion,os}$,  Mass of ionized gas lost from cloud due to photoevaporation \\
 $M_{\rm init}(<\thecs)$&Mass of initial cloud gas at $\theta<\thecs$ (Eq. \ref{eq:mej1})\\
 $M_{\rm ion}(t)$&$\int \dot M_\ionz dt +M_{\rm St,0}$, total mass of ionized cloud gas, not including gas ionized after it has left cloud (Eq. \ref{eq:mion})\\
  $M_{\rm ion}(>\thecs)$&Mass of ionized gas, including initial HII mass, produced at $\theta>\thecs$ (Eq. \ref{eq:miong} or for stall, \ref{eq:stall})\\
  $M_{\rm ion,os}$&Mass of ionized gas in the partially enclosed HII region between $A_o$ and $A_s$ (Eq. \ref{eq:mlossionz} or analytic Eq. \ref{eq:miios})\\
$M_{\rm loss}$&$ M_\ej+M_\evap$ (Eq. \ref{eq:mloss2}) or $M_{\rm ion}(>\thecs) + M_{\rm init}(<\thecs) - M_{\rm ion,os}$ (Eq. \ref{eq:mloss}), total mass lost from cloud \\
$M_{\rm St,0}$&Initial mass of HII region\\
 $M_\Omega$& Mass per unit solid angle centered on the association; $\delta M = M_\Omega \delta\Omega$; $M_{s,\Omega}$ is same in shell (Eq. \ref{eq:msom})\\
  $\nho$&Mean H density in cloud (Eq. \ref{eq:nh})\\
 $\nii(\theta)$&H nucleus density in HII region, assumed fully ionized (H$^+$)  along a ray at $\theta$ (Eq. \ref{eq:nii}) \\
 $P_s(R)$&Ambient cloud pressure at radius $R$; at cloud surface, $P_s=P_{\rm ISM}$ (Eq. \ref{eq:p0})\\
 $\pif$&$\rhoii(\cii^2+\vii^2)$, total pressure of ionized gas at the ionization front (Eq. \ref{eq:pif})\\
 $P_{\rm ISM}$&Total ISM pressure acting on cloud surface (thermal $P_{th}$  plus turbulent) (Eq. \ref{eq:pism})\\
 $r_c(\theta)$&Distance from association to cloud surface at angle $\theta$\\
 $\rcs$&Value of $r_c(\theta)$ at $\theta=\thecs$, where shell intersects surface of cloud; $r_{cs,f}$ is final maximum value at $\tion$\\ 
 $\rcz$&$r_c(\theta=0)$: minimum distance from association to cloud surface (Eq. \ref{eq:xircz})\\
 $\rcom$&Value of $\rcs$ at $150^\circ$ where the shell transitions from blister to cometary stage.\\
 $\rs(\theta)$&Radius of shell of neutral gas around HII region at an angle $\theta$ (Eq. \ref{eq:rhoii});
 $\rsf$ is final maximum value at $\tion$ \\
$\rst$& Str\"omgren radius (Eq. \ref{eq:rst}); $\rsto$ is the initial value (Eq. \ref{eq:rsto})\\
$R(\Theta)$&Distance from cloud center at angle $\Theta$; $R_s(\Theta)\,=$~distance to $r_s(\theta)$; $R_{cs}(\Theta_{cs})=$~distance to $\rcs(\thecs).$\\
 $\vecR_a$&Vector from initial cloud center to association; $R_c=R_a+\rcz$ (Eq. \ref{eq:xircz})\\
 $R_c$&Cloud radius (Eq. \ref{eq:rc})\\
 $R_{\rm gal}$&Distance to galactic center; at solar circle, $R_{\rm gal,0}=8.25$~kpc\\
 $S$& Ionizing photon luminosity of an association; $S_{49}=S/(10^{49}$~photons s\e).\\
 $S_{\rm bli}$&Critical value of $S$ that separates embedded stage from blister stage (Eq. \ref{eq:Sblinew})\\
$S_{\rm com}$&Critical value of $S$ that separates blister stage from cometary stage (Eq. \ref{eq:Scomfit})\\
 $S_{\rm flash}$&The value of $S$ that instantly creates HII region with $\rsto>\rcz$; no embedded stage (Eq. \ref{eq:flash})\\
 $S_{\max}$&$440\times 10^{49} \eamo M_6$ photons s\e, ionizing luminosity of association of maximum mass $M_{a,\max}=0.1\eamo M$\\
  $S_{\rm stall}$&The value of $S$ that produces pressure equilibrium at $\xi_{\rm stall}$ (Eq. \ref{eq:seq})\\
 $\tcom$&  $t_s(\xircom)$ (Eqs. \ref{eq:t}, \ref{eq:xircom}) Time for the shell to reach $\thecs=150^\circ\Rightarrow \xircs=\xircom$ \\
 $\tcomf$&${\min}(\tion, 2\tcom)$, final time for the cometary stage\\
 $\tff$&$(3\pi/32G\bar\rho)^{1/2}$, the free-fall time for gas of mean density $\bar\rho$\\
 $\tion$& Ionization-weighted lifetime of an association that fully samples the IMF; 4 Myr for $\sfn>10$ associations  (Eq. \ref{eq:tion}) \\
 $t_{\rm stall}$&$t_s(\xi_{\rm stall})$,  time for shell to reach its stall position (Eqs. \ref{eq:t} and \ref{eq:xieq})\\
 $\vesc$&Escape velocity from cloud (Eq. \ref{eq:vesc})\\
 $v_s(\theta)$&Expansion velocity of the shell at angle $\theta$ (Eq. \ref{eq:msomd}); analytic $v_s(t)$ (Eq. \ref{eq:vs})\\
 $\vii(\theta)$&Velocity of ionized gas flowing from the ionization front (Eq. \ref{eq:viitheta2})\\
 $\viieff$&A constant effective value of $\vii$ that provides analytic $M_{\rm loss}$ and $M_{\rm ion}$ matching numerical result
 (Eq. \ref{eq:viieff})\\
 $Z$&$R\cos\Theta$, distance along axis from cloud center; $Z_a$, to the association; $Z_o$, to $A_o$\\
 $\alpha_{\rm B}$&Case B hydrogen recombination coefficient = $2.59\times 10^{-13}$ cm$^3$ s\e\ at $T=10^4$ K \citep{drai11a}  \\
$\epsilon_{a,\max}$&$(M_{\rm a,max}/M)$, maximum star formation efficiency for a cloud to form an association; $\eamo=\epsilon_{a,\max}/0.1$\\
$\theta$&Angle between $\vecr$ and $\vecR_a$ (Fig. \ref{fig-3})\\
$\thecs$&The value of $\theta$ at which the shell intersects the cloud surface (Eq. \ref{eq:cos});
$\theta_{cs,f}$ is final maximum value at $\tion$\\
 $\Theta$&Angle between $\vecR$ and $\vecR_a$ (Fig. 1 and Appendix C); $\Theta_{cs}=$ angle between $\vecR_{cs}$ and $\vecR_a$\\
 $\mu$&$\cos\Theta$; $\mu_{cs}=\cos\Theta_{cs}$\\
 $\mu_r$&$\cos\theta$ (Eq. \ref{eq:mhii1}); $\murcs=\cos\thecs$ (Eq. \ref{eq:cos})\\
  $\xi$&$r/R_c$, normalized radius (Eq. \ref{eq:xi}); see definition of $r_x$ for different subscripts $x$ \\
  $\xi_\cen$&$R_a/R_c=1-\xi_{c0}$, normalized distance from cloud center to the association (Eq. \ref{eq:xircz})\\
  $\xircom$&Normalized distance from association to cloud surface at $\thecs=150^\circ$ (Eq. \ref{eq:xircom})\\
  $\xi_{\rm stall}(\thecs)$& Normalized shell radius for HII region in pressure equilibrium at angle $\thecs$ (Eq. \ref{eq:xieq})\\
$\rho_0$& Initial cloud mass density
($\rho_0=2.34\times 10^{-24}$ $\nho$ g cm$^{-3}$)\\
$\rhoii(\theta)$& Mass density of ionized gas along a ray at $\theta$ ($\rhoii=2.34\times 10^{-24}$ $\nii$ g cm$^{-3}$) (Eq. \ref{eq:rhoii}) \\
 $\Sigma$&$M/\pi R_c^2$, mean surface density of molecular cloud; $\Sigma_2=\Sigma/(10^2$\,\msun\ pc\ee)\\
 $\Sigma_s(\theta)$& $M_{s,\Omega}/r_s^2$, surface density of neutral shell along a ray from the association at angle $\theta$\\
 $\phiii$&$\pif/\rhoii\cii^2$, total pressure just behind the ionization front relative to thermal value (Eq. \ref{eq:phip})\\
 $\phiiieff$&Time-independent effective value of $\phiii$ needed for analytic solutions; $\phiiieff=1.2$ (Eq. \ref{eq:phiiieff})\\
 $\phieff$&Dimensionless parameter proportional to the
 stalling cloud pressure, derived by fit to $M_{\rm loss}$ and $\Sbli$ (Eqs. \ref{eq:Ps}; \ref{eq:phi-P-eff})\\ 
 $\phi_{\rm P,eff,bli}$&Value of $\phieff$ at $S=\Sbli$.
\end{longtable*}

Table 1 summarizes the symbols used in this paper.  To simplify notation, we suppress the
subscript ``cs" for the parameters 
$\xireq$, $\xircom$, and $\rcom$ 
 even though they refer to angles or distances from the association to where the shell intersects the surface.\\

\section{GMCs  and Young Ionizing Associations}
 \subsection{Cloud mass, radius, and average density}
 \label{sec:cloud}

 The properties of a spherical GMC with constant density  are completely specified by its
surface density, $\Sigma$, and its mass, $M$.  Given
these two parameters, its radius is 
\begin{equation}
   \Rc\equiv 56.4 \Sigma_2^{-1/2}  \Mcs^{1/2} \  {\rm pc},
   \label{eq:rc}
 \end{equation}
where $\Sigma_2\equiv \Sigma/(100$ M$_\odot$ pc$^{-2}$) and 
$\Mcs\equiv M/ (10^6$ M$_\odot)$.   Its hydrogen nucleus
density is
\begin{equation}
   \nho\equiv 38.4 \Sigma_2^{3/2}\Mcs^{-1/2} \ {\rm cm^{-3}}.
   \label{eq:nh}
 \end{equation}

 In some of our figures and results we focus on GMCs in the Milky Way.  Here,
 as discussed in Rosolowsky et al (in preparation), we use the $^{12}$CO($J=3-2$) observations of \citet{colo19} of GMCs in the Milky Way to identify GMCs and CO($J=1-0$) observations to derive the characteristic 
radius of a GMC, $\Rc=(A/\pi)^{1/2}$, where $A$ is the projected area of the cloud. 
In terms of  the \tco($J=1-0$) cloud mass, $M$, the  average  GMC radius depends on Galactocentric radius $R_{\rm gal}$ as (Rosolowky et al, in preparation)
 \begin{equation}
   \Rc=55 \Mcs^{0.46}10^{0.036 
  [R_{\rm gal}/(1\ {\rm kpc})-5.3]} \  {\rm pc}.
   \label{eq:rc1}
 \end{equation}
We take for our standard Milky Way case $R_{\rm gal}=5.3$
kpc, which is approximately the median radius of CO clouds between the Sun and the center of Galaxy.  The standard Milky Way case is then
$\Rc\simeq 55 \Mcs^{0.46}$\ pc. Equivalently, in
the Milky Way there is a relation between $\Sigma$
and $M$ such that just one of these parameters (e.g., the mass, $M$) is needed to specify the cloud properties.  The 
Milky Way relation is
\begin{equation}
     \Sigma_2=\frac{M/\pi R_c^2}{100\, \msunm\,{\rm pc}^{-2}}\simeq 1.05M_6^{0.08}.
     \label{eq:Sigma}
 \end{equation}
Thus, in the standard Milky Way case, the surface
 density is very insensitive to cloud mass,  in agreement with \citet{lars81} and \citet{roma10},  and is
 of the order 100 M$_\odot$ pc$^{-2}$.
 
 We emphasize that we present general results (clouds with any combination of $M$ and $\Sigma$) and as well as results specific to the Milky Way at $R_{\rm gal}\simeq 5.3$ kpc.   General results are
 needed since even in the Milky Way at a given
 $R_{\rm gal}$ there is significant dispersion around the average relation given in Equation (\ref{eq:Sigma}), and there is  variation of the mean $\Sigma$ with $R_{\rm gal}$.
 In addition, surface densities of GMCs vary significantly in external
 galaxies.

Although the results of this paper are general and are given as a function of cloud mass $M$ and surface density $\Sigma$,  we  focus here on a range of cloud masses from $10^4 - 10^7$ M$_\odot$ ($3\times 10^6$ M$_\odot$ for Milky Way clouds)   and surface densities 
$\Sigma_2=0.5-10$. 

\subsection{Stellar associations} 
\label{sec:evosc}

 \begin{figure}
\hbox{\includegraphics[scale=0.7,angle=0]{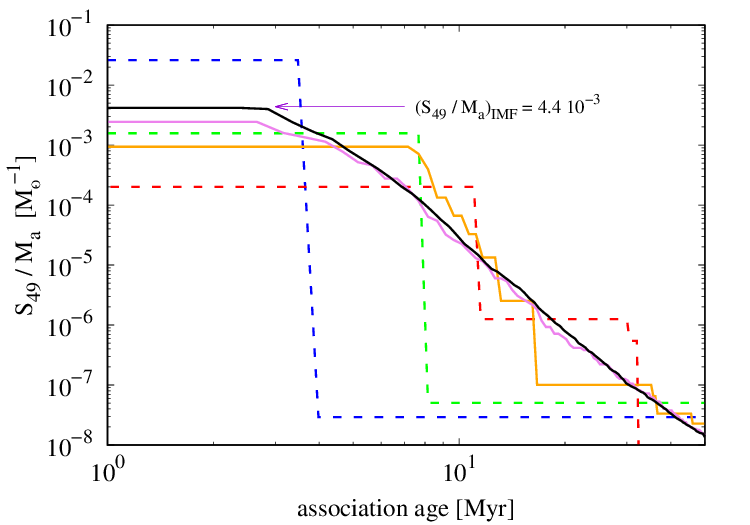}}
\caption{Evolution of the EUV  luminosity per unit stellar mass, $S_{49}(t)/M_a$,  for three associations with initial masses of 100 \msun (blue, green, and red dashed lines), one association of 1000 \msun (orange), one association of 10000 \msun (violet), and one association of 100000 \msun (black), which most fully samples the IMF.
The most massive star in the 100 \msun \ associations are 74 \msun  (blue), 12 \msun (green) and 5 \msun (red). The arrow indicates the initial EUV luminosity per unit stellar mass in a fully sampled IMF (see text). 
}

\label{fig-1} 
\end{figure}

  \begin{figure}
\caption{The EUV lifetime of an association as a function of $\sfn$.   The blue dots correspond to 1000 associations with masses between 40 and $10^5$ $\emsun$ following a power-law association mass distribution of index $-1$. Each association is created using the \citet{pg07} method for creating  a stellar mass distribution for an association of given mass.   The red points and dashed curve show $t_{\ionz}(S_{49})$ for single stars.
The diagonal black line is the fit $ t_{\ionz}= 6.05\, S_{49}^{-0.18}$ Myr to the average value of $t_{\ionz}$ of associations with $S_{49} < 10$. The horizontal black segment is the fit  $t_{\ionz}= 4$ Myr for associations with $S_{49} > 10$.
\hbox{\includegraphics[scale=0.7,angle=0]{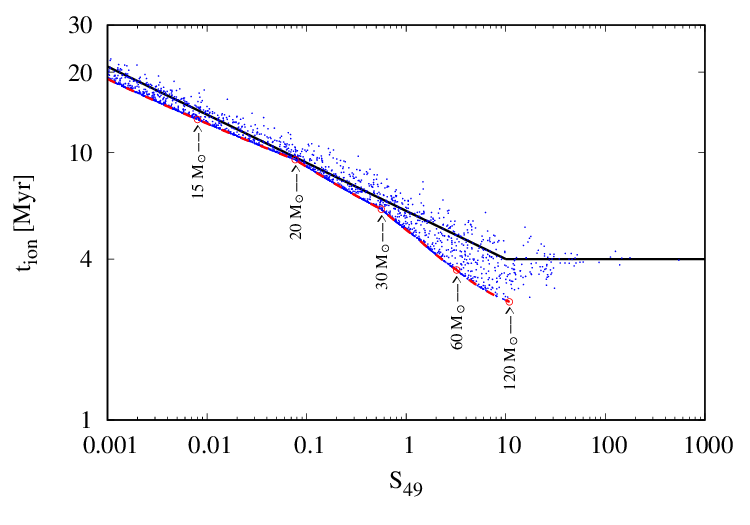}}
}
\label{fig-2a} 
\end{figure}

Within each cloud of mass $M$, we consider
a wide range of initially embedded associations with initial EUV luminosities 
  varying from a very low value
to a value, $S_{\rm max}$, corresponding to the largest possible association in such a cloud.  In terms of the star formation efficiency of a single association, $\epsilon_{a}$, the mass of an association in a GMC of mass $M$ is $M_{\rm a}=\epsilon_{a}M$.    For an association that fully samples the IMF,  the ionizing luminosities from \citet{parr03} and the IMF from \citet{parr11}\footnote{The ionizing luminosity produced by this IMF is intermediate between that used in Starburst99 \citep{leit99} and by \citet{murr10}.} imply that
\beq
\sfn=440M_{a,5}, 
\label{eq:s}
\eeq
where $M_{a,5}=M_a/(10^5$\,\msun), so that the maximum possible ionizing luminosity in a cloud of mass $M$ is 
\beq
S_{\rm max,49}= 440\eamo M_6.
\label{eq:smax}
\eeq
 We have normalized these quantities to $\eamo=\epsilon_{a,\max}/0.1$ and
$S_{\rm max,49}\equiv S_{\rm max}/(10^{49}$ s$^{-1})$.
We focus on $M_6>0.01$ so that $M_{\rm a,max}>10^3\eamo $~\msun; the assumption of a fully sampled IMF is accurate to within about 10 percent for such massive associations. However, we also treat the large number of smaller associations that do not
fully sample the IMF.
The normalization of 0.1 for $\epsilon_{a,\rm max}$ is consistent with the results of \citet{rei17} and \citet{how17}.  Note that $\epsilon_a$ is the SFE for a cloud to form a single association, and is less than the overall SFE of the GMC. 

The total ionizing (EUV) photon luminosity, $S(t)$, of an association of age $t$  is calculated numerically from a large number of stars  \citep{parr11}  with the ZAMS EUV photon luminosity $s(m)$ for a star of mass $m$ (in units of \msun), and main sequence lifetime
$\tms (m)$ given in \citet{parr03}.   These results are
very similar to those of \citet{ster03}.
Figure \ref{fig-1} shows the EUV time evolution for three representative mass associations from $1000-10^5$ M$_\odot$, as well as three separate simulations of the lowest mass (100 M$_\odot$) association. 
For a fully sampled IMF, the $t=0$ initial luminosity to stellar mass ratio is $\sfn/M_a=4.4\times 10^{-3}$, where $M_a$ is the total stellar mass in the association in solar masses.  In this paper $M_a$ does not enter as a parameter since we characterize a given association solely by its EUV luminosity, $S$.

  Figure \ref{fig-2a} shows
  how associations with $\sfn<10$ live longer on average than  associations with $\sfn>10$ because their EUV-producing stars are less massive.  These lower luminosity associations do not fully sample the stellar IMF and, in fact, their low ionizing luminosity demands that the highest mass star in the association is almost always less than the upper limit on stellar mass.   Typically, their ionizing luminosity is dominated by the
  single most massive star  and these relatively low-mass EUV-producing stars have much longer lifetimes than the $S$-weighted average lifetime of a fully sampled IMF seen for $\sfn>10$. A fit to the results plotted in Figure \ref{fig-2a} gives a median lifetime of an association of  ionizing luminosity $S$ of 
  \beq
  \tion\simeq \mbox{max}\left(\frac{6.05}{ \sfn^{0.18}},\; 4.0\right)~~~\mbox{Myr},
  \label{eq:tion}
  \eeq
  where $\tion$ is defined such that $S\tion$ is the total number of ionizing photons emitted by the association. 
  The crossover from the first term to the second occurs at  $\sfn=10$, corresponding to an association with a mass $M_a\simeq 2000$ ~\msun. 
For both our numerical and 
analytic solutions,
we assume constant $\sfn$ for $t<\tion$ and then a sudden drop to zero for $t>\tion$.

\section{Evolution of HII Regions}
\label{sec:evohii}

\subsection{Evolution of the shell and evaporated mass}
\label{sec:mass}

We now present the equations we use to follow the dynamics of the model HII region described in the previous section. As noted above, one of the key assumptions we make is that the motion of the gas is radial with respect to the association so that the solid angle subtended by an element of gas, $\delta\Omega$, is constant in time. This assumption would be valid if the HII region were centered in the molecular cloud. In fact, we assume that the HII region is off-center, so that this assumption becomes an approximation in our work.  We also assume that the shell is thin.

 The density of the ionized gas in the HII region governs the expansion of the HII region. In the numerical model, we assume that this density varies with angle but not radius and denote it $\rhoii(\theta)$.  (In the analytic solution, we assume that $\rhoii$ is independent of $\theta$.)  The value of the density is given by the Str\"omgren condition,
\beq
\rhoii(\theta)=\muh\left[\frac{3\fion S}{4\pi\alpha_{\rm B} r_s(\theta)^3}\right]^{1/2}.
\label{eq:rhoii}
\eeq
Here $r_s(\theta)$ is the distance from the association to the shell of neutral gas at an angle $\theta$, and $\fion S$ is the ionizing luminosity absorbed by the gas, not the dust. To simplify
notation, we generally drop the $\theta$, but it should
be understood that $\rhoii \equiv \rhoii(\theta)$ in the discussion of the numerical solution. Appendix A uses the results of 
\citet{drai11b} to determine $\fion$, which only varies from unity at low $\sfn n_{\rm II}$ to roughly 0.5 at high values.

 For a given value of $\nii$, Equation (\ref{eq:rhoii}) defines the Str\"omgren radius. The initial value of the Str\"omgren radius is  determined by the condition $\nii=\nho$,
\beq
\rsto=\left(\frac{3\fion S}{4\pi\ab \nho^2}\right)^{1/3}.
\label{eq:rsto}
\eeq
It follows that the density of ionized gas is given by
\beq
\nii=\left(\frac{\rsto}{r_s}\right)^{3/2}\nho.
\label{eq:nii}
\eeq

First we evaluate the mass of gas in the neutral shell that is expanding due to the HII region.
Let $M_\Omega=\delta M/\delta\Omega$ be the mass per unit solid angle; we term this the specific mass. In particular, the specific mass in the shell is $\msom$;
since the shell is thin, $\msom=r_s^2\Sigma_s$, where $\Sigma_s$  is the surface density of the shell. We define $M_{\rm ion}(t)$ as the total mass of gas that has been ionized  {\it inside} the cloud by time $t$ --i.e., excluding gas that is ionized after being ejected from the cloud.  
This mass is given by
\begin{equation}
    M_{\rm ion}= M_{\rm St,0} + \int^t_0 \dot M_{\rm ion} dt,
    \label{eq:mion}
\end{equation}
where $M_{\rm St,0}$ is the initial mass of the HII region and the second term is the integral of the mass flux through the ionization front of the shell while it lies inside the cloud.
Including the mass swept up from the ambient cloud, 
the ionized gas in the initial Str\"omgren region, and
the mass lost through
the ionization front, $\mionom$, the specific mass of the shell is
\begin{equation}
    \msom = \left[\frac 43 \pi \rho_0 r_s^3\right]\frac{1} {4 \pi}-\mionom
    \label{eq:msom}
\end{equation}
for $\rsto <r_s  <r_c$--i.e., so that the ionization front is still inside the cloud. In the case of $\thecs>150^\circ$, where part or all of the shell (i.e., the cometary cloud) lies beyond cloud surface ($r_s>r_c$), replace $r_s$ in above equation with $r_c$ since very little mass is swept into the shell beyond the cloud surface.  In the equation above, $\rho_0=\mu_{\rm H} n_{0}$ is the initial mass density of the cloud,  where $\muh=2.34\times 10^{-24}$~g is the mass per hydrogen nucleus. 
Since the flow rate per sterradian through the IF is
\beq
\frac{d\mionom}{dt}\equiv \dot M_{\rm ion,\Omega}=r_s^2\rhoii\vii,
\label{eq:mionomd}
\eeq
the rate of change of the specific shell mass as it expands in the cloud is
\beq
\frac{d\msom}{dt}=r_s^2(\rho_0v_s-\rhoii\vii),
\label{eq:msomd}
\eeq
In these equations, $v_s=dr_s/dt$ is the speed of the shell with respect to the association, and $\vii$ is the 
speed of the ionized gas behind the ionization front (IF) relative to the shell.

There are two different cases for the value of $\vii$. For an embedded HII region, the Str\"omgren condition gives $\nii^2r_s^3=$~const, so that $\mionom\propto\rhoii r_s^3\propto\nii r_s^3\propto r_s^{3/2}$ and
\beq
\frac{d\mionom}{dt}=\frac{3\mionom v_s}{2r_s}=\frac 12 r_s^2\rhoii v_s~~~~~\mbox{(embedded)}.
\label{eq:miiom}
\eeq
Equating this to the specific evaporation rate in Equation (\ref{eq:mionomd}) gives $\vii=\frac 12 v_s$. For embedded HII regions, the product $\rhoii \vii$ is the evaporated mass flux needed to fill the expanding HII region.
Generally, $\vii=\frac 12 v_s < \cii$ in embedded HII regions, so it is often neglected. 

After the HII region breaks out of the cloud and becomes a blister HII region, the value of $\vii$ increases as the area of the opening to the ISM, $A_o$, expands, and
therefore the pressure at the ionization front increases via the rocket effect.    
Appendix D provides the non-spherical solution
for $\vii(\theta)$  (Eq. \ref{eq:viitheta}) that we treat in our numerical solutions.
\beq
\vii(\theta)=\frac{\avg{\rhoiit v_s(\theta)}}{2\rhoii(\theta)}
+\frac{\rhoiics A_o}{\rhoiit A_s}\left(\cii-\frac 12 \dot r_{cs}\cos \theta_{cs}\right).
\label{eq:viitheta2}
\eeq
The bracket in the first term denotes an average over $\theta$.  

The flow of ionized gas from an ionization front in a partially enclosed  HII region is slower than in a D-critical ionization front ($\vii=\cii$) because the pressure in the HII region impedes the flow. 
Equation (\ref{eq:viitheta2}) is approximate
because if $v_s$ is large at $\theta>90^\circ$, $\vii$ can slightly exceed 
$\cii$, which is unphysical.\footnote{This occurs only for very large associations of mass $M_a\sim 0.1 M$, which are the only ones that can drive shells to $\theta>90^\circ$ at  high speeds.}   We therefore
restrict $\vii$ to the minimum of $\cii$ or to
the value given in Equation (\ref{eq:viitheta2}).

The mass loss rate to the ISM via evaporation, $\dot M_{\rm evap}$, is given by the rate at which gas flows out through the opening $A_o$. Let $M_{\rm ion,os}$ be the mass of ionized gas in the partially enclosed HII region, which is between $A_o$ and $A_s$ (see Fig. \ref{fig-3}). The mass loss rate to the ISM is then equal to the evaporation rate
from the shell in the cloud, $\dot M_{\rm ion}$, minus the rate of change of $M_{\rm ion,os}$, 
\begin{equation}
    \dot M_{\rm evap}= \dot M_{\rm ion}-\dot M_{\rm ion,os}.
    \label{eq:mlossionz}
\end{equation}
Appendix D provides solutions for 
$\dot M_{\rm ion}$ (Eq. \ref{eq:miond}) and $\dot M_{\rm ion,os}$ (Eq. \ref{eq:mhiid1}).  We find it convenient in the numerical analysis to then use Equation (\ref{eq:mlossionz}) above to solve for $\dot M_{\rm evap}$.

Finally, when the shell is outside the cloud ($r_s>r_c$), 
the back pressure due to the HII region decreases and we assume that in this case the flow off the surface is D-critical, with $\vii=\cii$.

\subsection{Equation of motion}
\label{sec:motion}

Next, we determine the equation of motion of the shell of neutral gas. The pressure at the ionization front, $\pif=\rhoii(\cii^2+\vii^2)$, 
accelerates the shell outwards, whereas the ambient cloud pressure $P_s$ and the ram pressure of the swept up ambient cloud gas (or ISM gas if $r_s>r_c$) decelerate the shell. 
We ignore radiation pressure and forces due to stellar winds (see the Introduction and Appendix B).
Momentum conservation for a mass element $\delta M_s$ in a solid angle $\delta \Omega$ then implies 
\begin{equation}
    \frac{d(\msom v_s)} {dt}=r_s^2 \rhoii[\cii^2+\vii(\vii-v_s)]-r_s^2P_s
    \label{eq:mom}
\end{equation}
(\citealp{matz02}, with the addition of an ambient pressure). 
With the aid of Equation (\ref{eq:msomd}), this leads to the equation of motion for the shell,
\beq
   \frac{dv_s}{dt}=  \frac{r_s^2}{\msom}\left( \pif-\rho_0 v_s^2-P_s\right).
   \label{eq:dvdt}
   \eeq
This is the equation we use in our numerical model for the evolution of HII regions. 

Note that outside the cloud, Equation (\ref{eq:dvdt}) applies with $\rho_0=\rho_\ism$. We take $n_{\rm ISM}=1$ cm$^{-3}$ or $\rho_\ism=2.34\times 10^{-24}$ gm cm$^{-3}$, appropriate to the WNM at $R_{\rm gal}\simeq 5$ kpc \citep{wolfire03}.  This ram pressure term is  small compared the cloud ram pressure term so that the shell accelerates on leaving the cloud until at much higher $v_s$ the ISM ram pressure can retard the acceleration.

\subsubsection{The rocket parameter, \texorpdfstring{$\phiii$}{phiII})}

We find it convenient to introduce the parameter
\beq
\phiii\equiv 1+\frac{\vii^2}{\cii^2}.
\label{eq:phip}
\eeq
so that
\beq
\pif=\rhoii( \cii^2+\vii^2)= \phiii\rhoii\cii^2.
\label{eq:pif}
\eeq
The parameter $\phiii$ measures the magnitude of the rocket effect and is key to the dynamics of the shell. 
We then have $\phiii\simeq 1+v_s^2/(4\cii^2)$ for an embedded HII region. For a blister HII region with $\theta_{cs}=90^\circ$,   $\phiii \simeq 1+ [\frac 12(v_s/\cii) +A_o/A_s]^2$; 
note that \citet{matz02} assumed that the ionization front in a blister was D-critical and adopted $\phiii=2$ for this case. Finally, we estimate
$\phiii=2$ for gas that has been ejected beyond the cloud since the ionized gas should be able to escape reasonably freely through the clumpy shell.  Therefore, 
$\phiii$ lies between 1 and $\sim 2$.

\subsubsection{External pressure, $P_s$} 

Under the assumption that the GMC was in hydrostatic equilibrium before the HII region formed,
the total pressure
inside the cloud needed to support it against gravity
and the pressure of the interstellar medium, $P_\ism$, is 
\begin{equation}
    P_s= \frac{GM\rho_0}{2R_c}\left(1 - \frac{R^2}{ R_c^2}\right) + P_{\ism}~~~~~(r<r_c).
    \label{eq:p0}
\end{equation}
Recall that $R$ is the distance from cloud center and $r$ is the distance from the association.
$P_s$ is a function of $r_s$  since $R$ is a function of $r_s$  (and $\theta$), but we have
suppressed the notation $P_s(r_s)$ for simplicity. We show in Appendix C that 
at our fiducial radius $R_{\rm gal}\simeq 5$~kpc in the Milky Way galaxy,  $P_\ism\simeq 3.7\times 10^{-12}$~dyne~cm\ee.

\subsection{Numerical Methods for Solution}

The numerical solution follows the dynamics of the shell
in 180 equally spaced angles in the range $\theta=0-180^\circ$, utilizing the equation of motion, Equation (\ref{eq:dvdt}).
Each of these 180 conical sections $i$ has fixed solid angle $\delta \Omega_i$. Each segment has a time-dependent shell mass $M_s(i,t)= \msom(i,t) \delta
\Omega_i$.  We follow the distance of each shell segment from the association, $r_s(i,t)$, the speed  of each shell segment, $v_s(i,t)$, and the mass of each shell segment, $M_s(i,t)$, (eqs. \ref{eq:msom}-\ref{eq:msomd}) as functions of time as the shell expands.

The equation of motion includes the cloud ambient pressure, $P_s$, which is a function of $R$ and therefore varies along each line of sight, both with $\theta$ and $r_s$.
This variation is not explicitly included in the analytic solution for the mass loss (see below subsection 5.3)  or in the standard analytic equations for the evolution of an HII region given in Section 5.
Thus, whereas the shell inside the cloud is spherical for the analytic solution, it is not spherical in the numerical one.

The numerical solution also follows the change in time of $\phiii$ as a shell evolves.   Note that $\phiii$ depends on both $r_s$ and $v_s$  (see Eq. \ref{eq:viitheta2}).   Therefore,  the equations of motion for the shell
segments and for the mass loss
can only be integrated numerically.
The analytic solutions use an effective
$\phiii=\phiiieff$ that is constant in time.

\subsection{Mass Loss}
\label{sec:massloss}

We treat the ejection of the neutral shell of gas across the cloud surface  due to the pressure of the expanding HII region as well as the  
 photoevaporation  of $\sim 10^4$ K
photoionized gas from the cloud into the ISM (sometimes called ``champagne flow"). 
Recall that we have ignored the effects of radiation pressure in accelerating the
neutral shell and in creating a positive gas density gradient in the HII region.
We have also ignored the retarding effects of gravity on the expanding shell and HII gas.  These limit the $\Sigma$, $M$ parameter space where our model is valid.
In Section 7 and Appendix B, we find that radiation pressure can be neglected as long as
\begin{equation}
    \sfn< S_{\rm ch,49}\simeq 1000\; \frac{M_6^{0.3}}{\Sigma_2},
    \label{eq:Sch}
\end{equation}
 corresponding to $M_a\la 2.3\times 10^5 (M_6^{0.3}/\Sigma_2)$~\msun. 

The gravity due to the cloud can retard the expansion of the
ionized gas.   In their studies of disk winds, \citet{bege83} and \citet{adams04} found that winds can arise even when the disk radius is significantly less than the gravitational radius, $R_g=GM/\cii^2$, corresponding to $\cii$ being less than the escape velocity, 
\begin{equation}
  \vesc= 12.4 \Sigma_2^{1/4}\Mcs^{1/4} \ {\rm km\  s^{-1}}.
  \label{eq:vesc}
\end{equation} 
For a disk, the kinetic energy is half the binding energy, making it easier to drive a wind. Adopting the conservative assumption that the kinetic energy of the ionized gas near the cloud is negligible, we find that 
the condition for gravity to restrict mass flow to the ISM is approximately $\cii \la 0.5\vesc$. This is consistent with the results of \citet{kim18}, who found that in one of their simulations more than 60\% of the cloud mass was driven away by photoevaporation in a cloud with $\cii=0.54\vesc$.  The condition that gravity not significantly affect our model, $\cii>0.5\vesc$ corresponds to
\begin{equation}
    \Sigma_2 M_6 \la 10.
    \label{eq:grav}
\end{equation}
In the rest of this section we assume these conditions are met and our model is valid.

\subsubsection{Ejection}    We distinguish ejection at $\thecs<150^\circ$  from the case in which $\thecs$ reaches beyond $150^\circ$ ($\xircs>\xircom$ for $\xircz=0.3$) and a cometary cloud forms. 
The shell ejected at 
$\theta>150^\circ$ becomes a
cometary cloud, and we follow the 
evaporation of the cometary cloud
as discussed below.  

{\it Cloud Mass Loss to ISM ($\thecs<150^\circ$}).  Numerically, at $\theta<150^\circ$ we find that if the shell makes it to the cloud surface, it is extremely likely to be accelerated to the escape speed $\vesc$. 
The acceleration is strong
because of the drop in retarding ram pressure of the ambient gas ahead of the shell, which is now ISM and not cloud gas.
In our numerical solution as each neutral shell segment $M_s(i)$ passes through the cloud surface, we add that mass to the total ejected neutral mass, $M_{\rm ej}(t)$. 
Once the shell exits the cloud and no longer 
sweeps up significant mass, the column density and thus $A_V$ through the shell diminish with time as the shell expands. 
We do not follow the
complicated photodissociation and photoionization of the shell after it is ejected from the cloud since here we are interested only in the mass lost from the cloud.

{\it Cometary Cloud Formation ($\thecs>150^\circ$}). Our numerical results show that by the time the shell has reached $\thecs\sim 150^\circ$, the shell has become approximately flat and perpendicular to the axis passing through association and cloud center.  As the shell moves beyond $\thecs\sim 150^\circ$ the shell becomes convex and is compressed into a cometary shape by the rocket effect and gravity.
Therefore, for $\thecs>150^\circ$ we consider all gas in the neutral shell  to be part of a cometary cloud.  However, we do allow for the photoevaporation of the
cometary cloud as discussed below.   In other words, ejected neutral gas at $\theta>150^\circ$ is not counted as mass loss, but for $\theta<150^\circ$ it is.
Ionized gas is mass loss in both cases, except that some ionized gas goes into the HII region for $\theta<150^\circ$; the "enclosed" (i.e., between $A_o$ and $A_s$) HII region is negligible for $\theta>150^\circ$. The dynamics are the same in both cases.  We note  that at $\thecs=150^\circ$, the initial cloud has already lost about 70\% of its initial mass by evaporation and ejection from $\theta<150^\circ$  so that the cometary cloud has at most about
30\% of the initial cloud mass.  Therefore, the cometary cloud regime applies only to high values of $S/M$ and results in
a narrow range of final total cumulative mass loss, $M_{\rm loss,f} \simeq (0.7-1.0)M$.

\subsubsection{ Photoevaporation, Supernova Ejection, and Total Mass Loss from Cloud}    

{\it Photoevaporation.} Numerically we find the total accumulated ionized mass produced from the sum of the angular segments.
This includes the initial HII region mass (see
Eq. \ref{eq:mion}).  We separate the photoevaporated gas to the ISM from the total ionized mass, which includes gas in the HII region (Eq. \ref{eq:mlossionz}). 
In addition,  high $\sfn$ associations  drive shells to $\thecs>150^\circ$ and the cometary cloud forms.  In analogy to WM97, we approximate this photoevaporative mass loss by following the evaporation of a shell with fixed solid angle set by $150^\circ < \theta <180^\circ$. We evaporate this expanding shell from $\tcom$ to $t_{\rm com,f}$, where $\tcom$ is the time for the shell to reach $\thecs=150^\circ$ and $t_{\rm com,f}$ is given by
\beq
t_{{\rm com},f}={\min}(2\tcom,\tion).
\label{eq:tcommax}
\eeq
As noted in WM97, this overestimates the evaporation from $\tcom$ to $t_{\rm com,f}$ because the cometary cloud has begun to form and compress so the illuminated area is less than assumed.  However, to compensate in the case where $2\tcom<\tion$ we assume zero evaporation after $2\tcom$ when, in reality, the evaporation of the cometary cloud continues to $\tion$. 

We have tested this simple approximation by also numerically following the evolution of a constant radius cometary cloud as it is driven from the association. The radius of the cloud is taken to be $1.54 R_c \sin 150^\circ=0.77 R_c$, the projected size at $\tcom$.  We find that the two approximations for $M_{\rm loss,f}$ match to better than 10\%
over the entire parameter range relevant to cometary clouds. 

{\it Effects of Supernovae.} A supernova has a relatively small effect on the destruction of its natal cloud: It will remove the ionized gas at $\theta<\thecs(t_{\rm SN})$ that is still attached to the cloud, but this mass is small and we ignore this ejection.
The supernova shock generally becomes radiative before reaching the edge of the Str\"omgren sphere for $\theta>\thecs(t_{\rm SN})$. After impact with shell/ambient cloud, the re-shocked gas cools, recombines and becomes part of the cloud, and therefore is not mass loss.

{\it Mass Loss from the Cloud}.   The total mass loss
from a cloud at time $t<\tion$ is then the sum of the evaporated ionized mass loss from the cloud plus the ejected
neutral mass loss
\begin{equation}
    M_{\rm loss}=  M_{\rm evap} +  M_{\rm ej} 
    \label{eq:mloss2}
\end{equation}
As noted above, if $\sfn$ drives the shell to $\thecs>150^\circ$, we do not count ejected mass at $\thecs>150^\circ$ as lost, but as a cometary cloud.\\

\begin{figure*}
\centering
\hspace{-0.4 cm}
\hbox{\includegraphics[scale=0.75,
angle=0]{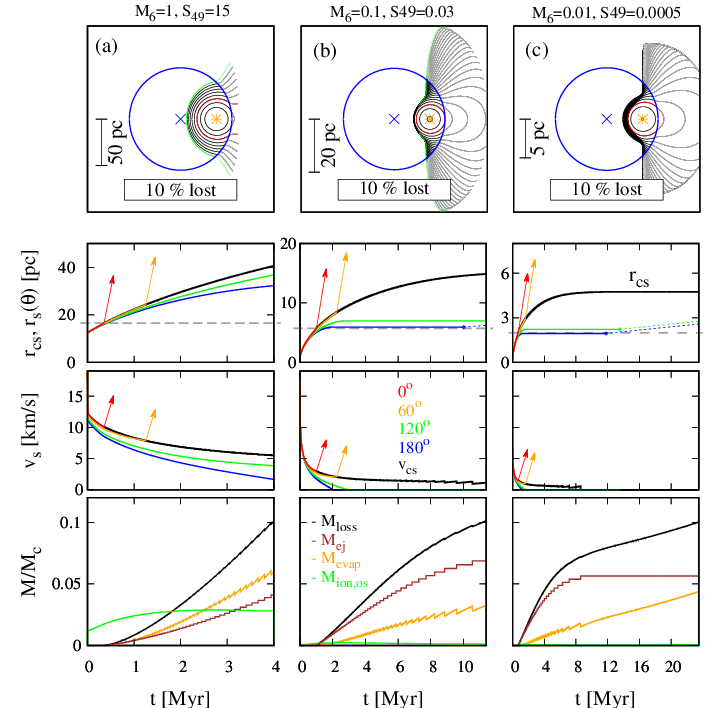}}
\hbox{\includegraphics[scale=0.75,
angle=0]{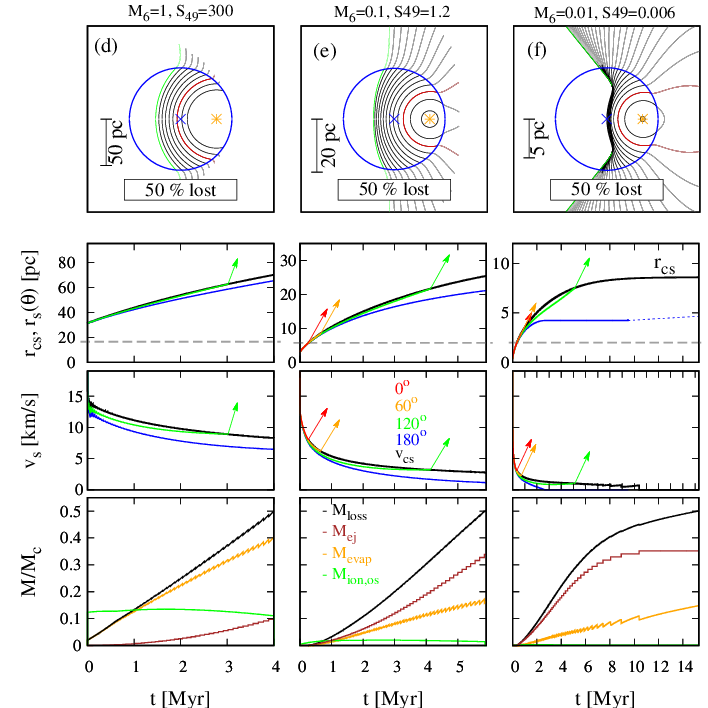}}
\caption{ The six top panels show the evolution of a shell around an association of ionizing luminosity $S_{49}$ at a depth $\xi_{c0}=0.3$ inside a cloud of mass $M_6=1, 0.1$ and $0.01$. The three left-hand panels, (a), (b) and (c), have luminosities  that produce a fractional mass loss $\sim 0.1$, whereas the three right-hand panels, (d), (e) and (f), have luminosities  that produce a fractional mass loss $\sim 0.5$.  We assume the Galactic surface density, $\Sigma_2=1.05 M_6^{0.08}$. The shell position is marked every $\Delta t=0.5$ Myr from $t=0$  to $\tion$. The red curve marks the position at $t=1$ Myr and the green curve at $\tion$ (4, 11.4, 23.5, 4, 5.9 and 15.5 Myr in the panels,  respectively). The blue circle is the initial cloud surface. When a curve in the region outside the initial cloud surface 
terminates, all the mass in the shell at that time and position has been evaporated.
The black curves in the second row panels show the evolution of the contact radius, $r_{cs}$, where the shell intersects the cloud surface. 
The red, orange, green and blue curves show respectively the evolution of $r_s$ in the $\theta=0$, $\theta=60^\circ$, $\theta=120$ and  $\theta=180^\circ$ directions. The colored diagonal arrows indicate the time when the shell in the corresponding direction crosses the cloud surface. If the shell in a given direction $\theta$ stalls and the shell gets completely ionized, the line representing $r_s(\theta)$ changes from solid to dotted and goes from horizontal to rising.  The increase of $r_s(\theta)$ after the shell gets completely ionized shows the advance of the ionization front into the ambient gas of the cloud; in the figure this only occurs in the 180° direction in case b, in the 120° and 180° directions in case c, and in the 180° direction in case f. The dash-gray horizontal lines in the second row show where $r=r_{c0}$; the red arrows show when the shell breaks out at $\theta=0$, so they originate at $r=r_{c0}$.
The third row panels show the evolution of the shell velocity in the four directions displayed in the second row panels. 
The bottom panels show the evolution of the accumulated total mass loss ($M_{\rm loss}$, black curve, Eq. \ref{eq:mloss2}), the neutral shell mass ejected from the cloud ($M_{\rm ej}$, brown curve), the ionized mass lost (evaporated) from the cloud  ($M_{\rm evap}$, orange curve) and the mass of embedded ionized gas ($M_{\rm ion,os}$, green curve). 
} 

\label{fig:shell} 
\end{figure*}

\subsection{Numerical Results of Shell Evolution}
\label{sec:results}

 Figures \ref{fig:shell}a, b, and c show the numerical evolution of a shell
whose driving association is located at $\xircz=0.3$ inside
 clouds of mass (a) $\Mcs=1$, (b) $\Mcs=0.1$, and (c) $\Mcs=0.01$.  In each case, $\sfn$ is chosen to produce a mass loss of 0.1$M$.    Figures \ref{fig:shell}d, e, and f are identical, except the $\sfn$ values have been increased so that the mass loss is 0.5$M$. 
 These three cases are in the blister state, but close to 
 entering the cometary cloud category, which occurs when
 $M_{\rm loss}\sim 0.7 M$.  In all cases in the figure, the Milky Way relation
 between $\Sigma_2$ and $M_6$ is used
 (Eq. \ref{eq:Sigma}).  If contours end, it indicates the shell has completely evaporated.
 
As seen in the top row, the expanding HII region is nearly spherical at early times during the embedded stage.   Once the shell breaks out of the cloud, the champagne flow is at first quite collimated, but the mass-loss cone widens with time until the shell stalls in the cloud (cases b, c and f) or the association dies at $\tion$ (cases a, d, and e).  The shell inside the cloud is then less spherical and more flattened, especially in
 cases where $\sfn$ is larger and the shell penetrates to and beyond
the cloud center. The flattening is due to the higher pressure, $P_s$, near the cloud center.  In the stalled cases--(b), (c) and (f)--one sees that there is a critical $\theta_{\rm cr}$ where, for $\theta< \theta_{\rm cr}$, the shell does not stall but proceeds to the cloud surface and beyond.  This critical point occurs inside the cloud and is due to the interplay of the decrease of cloud pressure with $R$ and the decrease of HII driving pressure with $r_s$.   For
$\theta< \theta_{\rm cr}$, the HII pressure is enough to push
the shell over the ``hill" of
the cloud pressure.
In addition, it is noteworthy that the final opening angle, $\theta_{\rm cs,f}$, is nearly the same in each fractional mass loss case;
$\theta_{\rm cs,f}\simeq 92^\circ$ for
$M_{\rm loss,f}/M=0.1$ and $\simeq 130^\circ$
for $M_{\rm loss,f}/M=0.5$.   Unless the shell stalls long before $\tion$, mass loss closely relates to $\theta_{\rm cs,f}$ because much of the mass loss is from the  evaporated and ejected mass in the loss cone $\theta<\theta_{\rm cs,f}$.   However, if the shell stalls with $t_{\rm stall}\ll \tion$, then for
$t_{\rm stall}<t<\tion$ the cloud loses mass by evaporation from $\theta>\theta_{\rm cs,f}$ 
and the final mass loss does not correlate as well with the size of the loss cone.

The second row of Figure \ref{fig:shell} shows the time
evolution of $r_s$ in various 
$\theta$ directions and also the 
increase of $\rcs$ with time. The arrows indicate ejection of
that portion of the shell into the ISM.  Note that $r_s(180^\circ)$
is smaller than $r_s$ at smaller
angles because of the higher cloud pressure encountered.  This flattening is most extreme
in case (f) where the shell stalls but has sufficiently high $\sfn$ to drive the shell at
180$^\circ$ to the cloud center,
where the cloud pressure peaks.  
As noted in the caption, the $r_c(\theta)$ curves become dashed if the shell completely evaporates as seen in the stalled shells at large $\theta$ in panels b, c and e. The EUV from the star then eats into the ambient cloud and the shell slightly advances with time.  For our analytic approximate solution
(discussed in next section and Appendix E) we find we can assume shell stalled  from $t_{\rm stall}$ to $\tion$ analytically and still get good agreement for the mass loss.

The third row shows the evolution of $v_s$ at various angles as well as the evolution
of $v_{\rm cs}$.  Note that for smaller $M_6$, the gas is denser, so the shell decelerates much more rapidly.  In cases (a),
(d) and (e) the shell has not stalled but is still moving at
$\tion$, when the association
``dies" (that is, $\sfn$ very rapidly decreases).
We see that the shell almost always
is traveling at less than $\vesc$  when it reaches the
surface.   However, we find
 it then rapidly accelerates
to escape speed because of the drop in the ram pressure as the shell passes from cloud  to ISM.

The last row is the most significant for this paper, which focuses on mass loss from clouds.  The black curve, $M_{\rm loss}$ is the sum of the
red curve (neutral shell mass ejected) and the orange curve (mass evaporated).   Of note is that for lower mass clouds ($M_6\la 0.1$), the
mass loss due to ejection  of neutral gas is greater than the evaporative mass loss.  The green curve represents HII mass lying between $A_o$ and $A_s$, which
recombines after $\tion$ and is not counted as mass loss.

Figure 5 is the same as Figure 4, except that the maximum $S_{\rm max,49}$ is adopted for each cloud mass case.  For $M_6=0.1$ and 0.01, this results in a cometary cloud at the end of the evolution.   However, even with $S_{\rm max,49}$, the $M_6=1$ cloud never reaches the cometary stage.
In the top row, one sees that for $M_6=0.1$ and 0.01 the shell transitions in shape from concave (as seen from association) to  convex as the shell (the cometary cloud)
emerges from cloud,  due to the higher ambient cloud pressures and larger surface densities in the shell at larger $\theta$.    The shell contours end when the shell totally evaporates and the figure shows that this happens rapidly beyond the cloud boundary at small $\theta$, where less cloud mass has been swept into the shell.  Gravity will also collapse the shell toward the
$\theta=180^\circ$ axis, but we have ignored this over our relatively short evaporation interval of the smaller of $\tion$ and $2\tcom$.  The figure shows that even with the maximum $\sfn$ for a given cloud mass, the cloud is not totally destroyed by a single association.   In the low cloud-mass cometary cases, some cometary cloud mass persists ($<0.1M$)  after the association dies. In the $M_6=1$ case, $\sim 40$\% of the initial cloud survives.  The bottom row shows that evaporation dominates cloud mass loss over $\theta<150^\circ$ ejection in all three mass cases with $\sfn=S_{\rm max,49}$.   However, the ratio of ejected mass to evaporated mass increases as the cloud mass decreases.

\begin{figure}
\hbox{\includegraphics[scale=0.70,
angle=0]{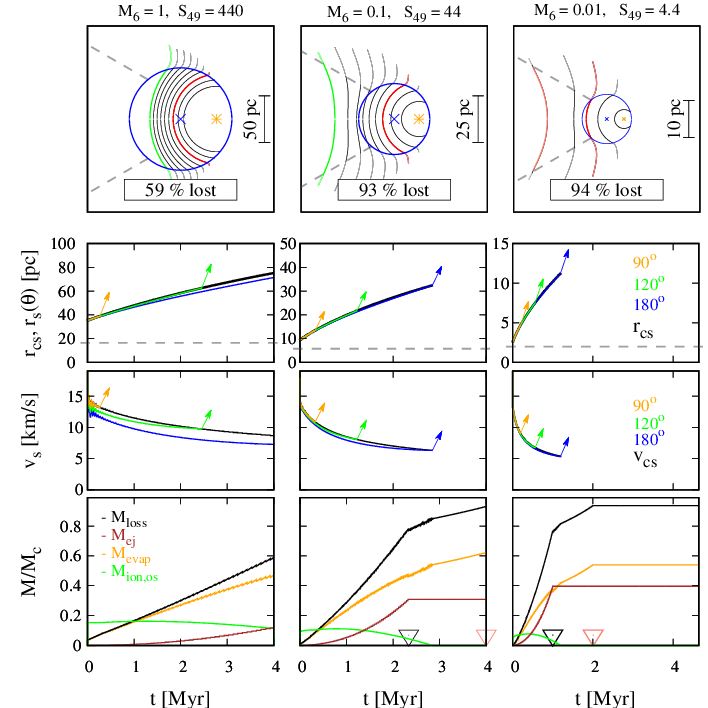}}
\caption{ Same as Figure \ref{fig:shell} but for $S_{49}=S_{{\max},49}=440 M_6$. The shells do not stall. The center and right columns show cases where the final state is cometary. The dash-gray lines drawn in the top row at $\theta=150^\circ$ show the portion of the shell after ejection from a cloud that is considered cometary.
The cometary stage is not attained in the case $M_6=1$ and $S=S_{\max}$, nor is it achieved in any of the cases shown in Figure \ref{fig:shell}.  
The black and salmon triangles in the bottom panels indicate the times $\tcom$ and $t_{\rm com,max}$, respectively (see text). 
} 
\label{fig:shell-Smax} 
\end{figure}

Finally, we compare our numerical results with the high resolution 3D hydrodynamical study of an O7 star placed at the center of  clumpy $10^4$ M$_\odot$ clouds of radius 6.4 pc or $\Sigma_2=0.78$ \citep{walc12}.  These authors study clouds with various fractal dimensions ranging from 2 to 2.8, which span large clumps to small clumps.  They find small differences in the evolution of the mean distance to the ionization front as a function of fractal dimension, and our ionization front evolution closely matches theirs to 10\% while the shell lies inside the cloud.
Their star is at cloud center, or $\xircz=1$, but as we have noted, we use smaller $\xircz$ to simulate clumpiness 
since clumpiness indicates lower column densities to the surface on some lines of sight. 
They define dispersal as material passing through the initial cloud radius and find complete dispersal of the cloud in 1-2 Myr (although at least half of the cloud
disperses in $\sim 1$ Myr in most of their models according to their Figure 6), whereas we find times for our shell to completely emerge from the cloud of 1.5, 1.1, and 0.9 Myr
for $\xircz$= 0.5, 0.75, and 0.9.  We both find that the total mass loss is dominated by mass loss at late times, $t\sim 0.5-1.5$ Myr, where our mass loss rates are comparable to theirs.
Finally, they find that the ratio at late times of the ejected neutral mass to the ionized evaporated mass ranges from about 3 to 8, depending on fractal dimension.   We find
that in our models we get ratios of 2.5, 3.8, and 5.3
for $\xircz$= 0.5, 0.75, and 0.9.  We conclude that our simple constant density cloud models provide a  good approximation for mass losses from clumpy clouds, a conclusion also found by \citet{krum06}.

We turn now to approximate analytic solutions that will provide insight into the numerical results we present in sections 6 and 7.

\section{Analytic Approximations to Evolution of HII Regions}
\label{sec:anal}

We have several motivations for finding approximate
analytic approximations to the dynamics and 
accumulated mass loss: They show the  the dependence on
$\Mcs$, $\sfn$,
$\xircz$, and $\Sigma_2$ for many of the dynamical parameters; 
they give approximate solutions
for any set of these parameters, beyond the numerical solutions we present; and they provide a check on the numerical code.  Finally, they reduce the computational time in  large simulations that compute the lifetimes of GMCs
due to EUV destruction.

In order to integrate the momentum Equation (Eq. \ref{eq:mom}) and the equation of motion (Eq. \ref{eq:dvdt}), we neglect the cloud pressure, $P_s$, and replace $\vii(t)$ and $\phiii(t)$ with constants, $\viieff$ and $\phiiieff$. 
These two effective parameters approximately
account for the time dependence of  $\phiii$ and $\vii$.
Initially,
the shell is embedded or nearly embedded
and $\vii \ll\cii$ and $\phiii\sim 1$ (see Eq. \ref{eq:vii}).  However, as time evolves the shell expands, the area of the opening to the ISM grows, and
$\vii$ and $\phiii$ grow.  If the association is large so that the shell grows to $\xirs>1$, then $\vii \sim \cii$
and $\phiii\sim 2$.   Fixing these two parameters as constant
in order to analytically solve the equation of motion is equivalent to using weighted average values throughout the time
evolution. For
$\viieff$, we take the geometric mean of the minimum value of $\vii$
at $t=0$ and its maximum value at $t=\tion$ (see Appendix E and discussion of Eq. \ref{eq:viieff} for $\viieff$). 
For $\phiiieff$ we also take the mean of the minimum value=1 of $\phi_{\rm II}$ and its maximum value=2, or
\begin{equation}
    \phiiieff=2^{1/2}
    \label{eq:phiiieff}
\end{equation}
and note that $\phiiieff$ has only a weak effect on $r_s$ and $v_s$ (Section \ref{sec:spitzer}). 

Although we neglect the cloud pressure in integrating the equation of motion and the momentum equation,
we must include it when we compute when and where the shell stalls.  In the absence of the cloud pressure, the dynamics are independent of $\theta$. In keeping with that approximation, 
we use an
average value for the cloud pressure, $\bar P_s(R_s)$,  and then set the HII driving pressure equal to that to determine when and where the shell stalls.
This average pressure can be expressed in terms of a dimensionless parameter, $\phieff$.  This parameter is discussed in Section 5.1.2 below, where we present an analytic equation for it that fits both the condition for the shell to break out of the cloud and the final accumulated mass loss $M_{\rm loss,f}$.
In essence, our analytic ``physical fit" to
the numerical model for $M_{\rm loss}$ as a function of $t$, $S$, $\Sigma$, $M$, and $\xircz$ just requires a single dimensionless fitting  function, $\phieff(S,\Sigma, M, \xircz)$, plus two parameters, $\viieff$ and $\phiiieff$.

In the rest of this section, we use $\phiiieff$, $\phieff$, and $\viieff$ to
solve for $r_s(t)$ and $v_s(t)$, to solve
for the stall criterion, 
and to derive analytic approximations for critical values of $\sfn$ that delineate different stages of evolution of an HII region.
Using these effective parameters, we present  a prescription for  the
analytic ``physical fit" solution for
$M_{\rm loss}$ in Appendix \ref{app:analytic}.\\

\subsection{Dynamics}
\subsubsection{Modified Spitzer Model for the expansion of an HII region}
\label{sec:spitzer}

 The radius of an HII region is given by the Str\"omgren relation (cf. Eq. \ref{eq:rhoii}),
\beq
\rst=\left(\frac{3\fion S}{4\pi\alpha_{\rm B}\nii^2}\right)^{1/3},
\label{eq:rst}
\eeq
where $\nii$ is the number density of ionized hydrogen in the HII region, which we assume to be  independent of position for ionized gas inside the cloud. 
Recall that
$\fion S$ is the ionizing luminosity absorbed by the gas, not the dust. The Case B recombination coefficient for hydrogen 
 at $T=10^4$~K, the temperature we adopt, is
$\alpha_{\rm B}=2.59\times 10^{-13}$ cm$^3$~s\e\ \citep{drai11a}. 
 Numerically, the initial value of the Str\"omgren radius is (Eq. \ref{eq:rsto})
\beqa
\rsto&=&3.15\left(\frac{\fion \sfn}{\nhot^2}\right)^{1/3}~\pc,\\
&=&6.0 \left[\frac{(\fion \sfn  \Mcs)^{1/3}}{\Sigma_2}\right]~\pc,
\label{eq:rsto1}
\eeqa
where $\nho$ is the constant initial hydrogen nucleus density of the cloud and
$\nhot=\nho/(100$ cm$^{-3})$ is given by Equation (\ref{eq:nh}).
When normalized to the cloud radius, this becomes
\beq
\xirsto=\frac{\rsto}{R_c}=0.106\, \frac{(\fion \sfn)^{1/3}}{\Sigma_2^{1/2}\Mcs^{1/6}}.
\label{eq:xisto}
\eeq
The density of gas in the HII region 
and therefore the  thermal pressure driving the expansion drops
as $r_s^{-3/2}$ as the HII region expands
(see Eq. \ref{eq:rst}).

The expanding ionized gas drives a shock into the surrounding cloud, which produces an expanding shell of dense, neutral gas. 
We approximate the equation of motion for the shell, Equation (\ref{eq:dvdt}), by setting the pressure behind the ionization front equal to $\pif=\phiiieff \rhoii\cii^2$, with $\phiiieff=2^{1/2}$, and by setting $P_s=0$. Integration then gives  
\beq
r_s=\rsto\left[1+\frac{7}{4}\left(\frac{4\phiiieff}{3}\right)^{1/2}\frac{\cii t}{\rsto}\right]^{4/7},
\label{eq:rs}
\eeq
where $\cii=11.1\,T_4^{1/2}$~km~s\e\ is the isothermal sound speed of the ionized gas.
The factor $(4\phiiieff/3)^{1/2}$ differs from the classical \citet{spit78} solution.  The factor $(4/3)^{1/2}$ was given by \citet{matz02} and Equation (\ref{eq:rs}) with $\phiiieff=1$ was given by \citet{hoso06}. The factor $(4/3)^{1/2}$ allows for the pressure drop that decelerates the shell.
The factor $\phiiieff$ allows for  the rocket effect \citep{matz02}.

The velocity of the expanding shell is
\beq
v_s=\left(\frac{4\phiiieff}{3}\right)^{1/2}\left(\frac{\rsto}{r_s}\right)^{3/4}\cii
\label{eq:vs}
\eeq
for $r_s\geq\rsto$ from Equation (\ref{eq:rs}).  The age of the shell  is
\beqa
t_{s6}&=&13.2 \frac{\Sigma_2^{-1/8}\Mcs^{5/8}}{\phiiieff^{1/2}(\fion \sfn)^{1/4}}\nonumber \\
&&\times\left[1-\left(\frac{0.106\fion^{1/3} \sfn^{1/3}}{\Sigma_2^{1/2}\Mcs^{1/6}\xi_s}\right)^{7/4}\right]\xi_s^{7/4}
\label{eq:t}
\eeqa
from Equations (\ref{eq:nh}) and (\ref{eq:rs}),
where  $t_{s6}$ is the time in Myr.
Alternatively, 
this equation implies
\beqa
    \xirs(t)= 0.23 \, \phiiieff\,^{2/7}\Sigma_2^{1/14}\Mcs^{-5/14} (\fion \sfn) ^{1/7}
\nonumber\\
&&\hspace{-6.5 cm}
    \times \left(1+\frac{0.26\Sigma_2^{-1}\Mcs^{1/3} \fion^{1/3}\sfn^{1/3}}
    {\phiiieff\,^{1/2} t_{s6}}\right)^{4/7} t_{s6}^{4/7} 
    \label{eq:xis}
\eeqa
 The second term in parenthesis is only important at early times when $\xirs \sim \xirsto$; it increases $\xi_s$ by a factor $\la 1.2$ for $\xi_s>2\xirsto$. The parameters
 $v_s$, $r_s$, and $\xi_s$ depend on a weak power (2/7) of $\phiiieff$.
 Therefore, below we simply substitute our effective value
 $\phiiieff=2^{1/2}$ into all equations.


\subsubsection{Stalled HII regions}
\label{sec:stall}

The expansion of the HII region can be halted by the pressure of the ambient cloud,  $P_s$; we term this a ``stalled  HII region''. In the numerical work, because $P_s$ depends on $R$, this causes the stall to occur at smaller distances from the association as $\theta$ increases.  We approximate the
$R$-dependent $P_s$ with 
a constant average value, $\bar P_s$, representative of stalling at $\thecs$, as suggested by our
numerical results.   We introduce the dimensionless
parameter $\phieff(S,\Sigma,M,\xircz)$ that determines the magnitude of this pressure (see Eq. \ref{eq:p0}),
\beq
    \bar P_s\equiv \frac{\phieff GM\rho_0}{R_c} + P_{\rm ISM}.
    \label{eq:Ps}
\eeq
We find that for low $S$, so that the shell at $\tion$ has just broken out of the cloud, $0.025\la\phieff\la 0.1$, but for high $S$, so that the shell at $\tion$ lies at high $\thecs$ and near cloud center, $\phieff$ can approach unity, as expected. We normalize the equations below to a value 0.1. An explicit expression for $\phieff$ is given in Equations (\ref{eq:phi-P-eff-bli}) and (\ref{eq:phi-P-eff}) below. 
 
The above expression for $\bar P_s$ can be put in terms of the surface density of the cloud,
\beqa
    \bar P_s&=& 0.24 \left(\frac{\phieff}{0.1}\right)G \Sigma^2 + P_{\rm ISM},\nonumber\\
    &\equiv&  \left[0.24 \left(\frac{\phieff}{0.1}\right) G \Sigma^2\right]C,
    \label{eq:bPs}
\eeqa
where the numerical value of $C$ is 
\begin{equation}
    C=1 +0.53 \left(\frac{0.1}{\phieff}\right)\left(\frac{P_{\rm ISM}}{3.7\times 10^{-12}\ {\rm dyne\ cm^{-2}}}\right) \Sigma_2^{-2}  .
\label{eq:C}
\end{equation}

The thermal pressure inside the HII region is
$\pii=2.1\nii k T_{\rm II}$ for fully ionized H and neutral He with an abundance 0.1 relative to H.
The stalling condition is $\pif=\phiiieff \pii=\bar P_s$. 
Inserting $\nii$ from Equation (\ref{eq:nii})  into $\pii$, we find that the value of the ionizing luminosity that results in an HII region that stalls at the surface at a normalized radius $\xi_{cs}$ is given by
\begin{equation}
   \Seqfn(\xi_{cs})=\frac{1.7}{\fion} \left(\frac{\phieff}{0.1}\right)^2C^2 \Sigma_2^{5/2} \Mcs^{3/2}\xi_{cs}^3
\label{eq:seq}
\end{equation}
This equation can be inverted to express the normalized stall radius $\xireq$  as a function
of the luminosity $\sfn$ of the association in the cloud,
\begin{equation}
    \xireq=0.85\left(\frac{0.1}{\phieff}\right)^{2/3}\frac{(C^{-2}\fion\sfn)^{1/3}}{\Sigma_2^{5/6} M_6^{1/2}}.
    \label{eq:xieq}
\end{equation}
Since the shell in our analytic work is spherical, this is the value of $\xircs$, the normalized radius to the intersection of shell and cloud surface; Equation (\ref{eq:cos})  then provides $\thecs$ for a stalled shell. As we show below, $\phieff$ is a function of $\sfn$, so Equation (\ref{eq:seq}) must be
modified to be non-transcendental.

{\it $t_{stall}$ and $t_{com}$}.  The time for an expanding shell to reach the stall point, $\teq$,
can be found by substituting the above
expression for $\xireq$ into Equation (\ref{eq:t}).   The time $\tcom$ to reach the cometary stage ($\thecs=150^\circ$ or $\xircs=\xircom$) is given by inserting
$\xircom$ into Equation (\ref{eq:t}), i.e., $\tcom=  t_s(\xircom)$.  

  \begin{figure*}
  \centering
\hbox{\includegraphics[scale=1.4,
angle=0]{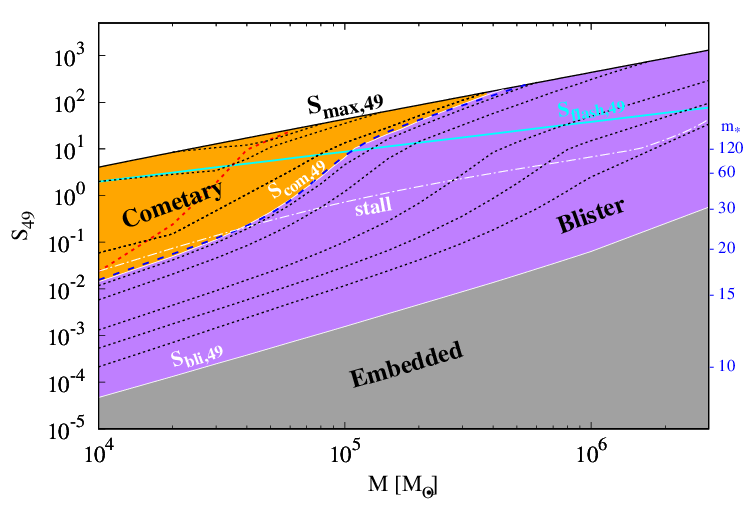}}
\caption{ For the Milky Way relation of $\Sigma$ to $M$ (Eq. \ref{eq:Sigma}) and for an association position $\xircz=0.3$, the figure shows as a function of $M$ 
the various categories in which shells driven by $\Sc$ end up at $t_\ionz$.    All associations with $S<\Sbli$ stay embedded, but those with $S>\Sbli$ break out of the cloud and drive mass loss via
partial ejection  of neutral gas  and  photoevaporation of ionized gas. 
 Between $\Sbli$ and $\Sflash$ the HII region is initially embedded but expands and breaks out of the cloud.  For $\sfn>\Sflash$, the HII region starts in the blister stage.  Between $\Sbli$ and $\Scom$ the shell either stalls or the association dies in the blister stage.  The cometary cloud stage is the final state between $\Scom$ and $S_{\max}$, which scales as the maximum star formation efficiency for a single association, assumed to be 0.1 in the figure.  From bottom to top, the black-dashed curves correspond to fractional mass losses of 0.05, 0.1, 0.2, 0.5, 0.7, 0.8, 0.9 and 0.95. Clouds with masses $M_6<0.4$ and high $\sfn$ end their evolution as small cometary clouds; clouds with masses $M_6>0.4$ never reach the cometary phase even with $S_{\max}$. 
The dashed red curve that separates the cometary region into two parts satisfies the condition  $2 \, t_{\rm com}=t_{\rm ion}$. To the right of this curve, $\tion<2\tcom$, and to the left, $2\tcom<\tion$; in the latter case  we stop cometary cloud evaporation at $2\tcom$ in the numerical code (see text).  The blue-dash curve that runs close to the numeric results for $S_{\rm com,49}$ is the solution of the parametric fit given in Equation (\ref{eq:Scomfit}) assuming the Milky Way relation of $\Sigma$ to $M$. The white dash-dot curve indicates the $S_{49}$ values below which the shell stalls in the direction $\theta=180^\circ$ before $t_{\rm ion}$. 
Labeled on right are
the average $\sfn$ of individual stars of different mass, but recall that $\sfn$ on the left is the sum of the ionizing luminosities of all the stars in an association.
}
\label{fig-6} 
\end{figure*}
 
\subsection{Evolution Categories and Critical Values of $\sfn$}
\label{sec:cat}

In order to evaluate the 
mass loss as the HII region
evolves, we consider three different categories of HII region: embedded, blister, and cometary cloud. These categories are shown in Figure \ref{fig-6} for the Milky Way ($\Sigma_2=1.05M_6^{0.08}$) for the case $\xircz=0.3$.   We identify the
categories in terms of the final state, denoted by the subscript ``$f$" in the text below, that a shell driven by an association with
given $\Sc$ reaches at $t_\ionz$. In this section, we adopt $\tion$ from Equation (\ref{eq:tion}). 
Figure \ref{fig-6} shows that for a given cloud mass, $M$, the categories represent a sequence in $\sfn$.   As $\sfn$ increases, the final state of the association/shell evolution goes from
embedded, to blister, and finally 
(for $M_6< 0.4$) 
to cometary cloud. 
 The critical values of $\sfn$ that mark the transition to each of these stages are
 $\Sbli$ and  $\Scom$.

These categories also mark a sequence in time for luminous associations and therefore stages in the evolution of an HII region. For example, a luminous association with $\Scom<S<S_{\rm max}$ will pass in time through embedded, blister and cometary stages.   Only HII regions with low $S<\Sbli$ are in the embedded category for their entire lifetime. All HII regions whose shells expand beyond $\xircs>\xircz$
 ($S>\Sbli$)  pass through the embedded stage. Very high luminosity associations ($S>\Sflash$) produce an initial HII region with $\xirsto>\xircz$ and are never embedded, but instantly become blister HII regions, and from there may evolve to the cometary stage.  
$S_{\rm max}$, the maximum possible luminosity of an association in a cloud of mass $M$, is given by Equation (\ref{eq:smax}).

Provided that the HII region reaches the blister stage, whether the shell stalls or becomes a cometary cloud depends on how $t_{\rm stall}$ and $\tcom$ are related to $\tion$: If $\tion$ is the shortest time, so that the association dies first, then the HII region is an expanding blister throughout its life; if $t_{\rm stall}$ is the shortest, then the HII region evolves from an expanding blister to a stalled blister (below the line labeled ``stall" in Fig. \ref{fig-6}); and if $\tcom$ is shortest, then the HII region evolves from an expanding blister to a cometary cloud.

\subsubsection{Embedded HII region ($\xi_{cs,f}<\xircz$, $S<S_{\rm bli}$)}

Embedded HII regions are contained within the natal molecular cloud for their lifetimes.
For embedded HII regions, the pressure of the cloud, $P_s(R)$,  stalls the shell before the shell reaches the surface in the $\theta=0^\circ$ direction. 
The condition for an HII region to be in the embedded category-- i.e., to have not entered the blister stage-- is  that the HII region stall before reaching the surface, so that $\sfn<\Seqfn(\xircz)\equiv \Sblifn$, where
\beqa
\hspace{-1cm} S_{\bli,49} &=&\frac{0.10}{f_{\rm ion,bli}} \left(\frac {\phi_{\rm P,eff,bli}}{0.1}\right)^2\times \nonumber \\&&~~~~~ \left(\frac {C_{\bli}}{1.5 }\right)^2 
\left(\frac{\xi_{c0}}{0.3}\right)^3\Sigma_2^{5/2}M_6^{3/2},
\label{eq:Sblinew}
\eeqa
 from Equation (\ref{eq:seq}) 
and where the subscript ``bli" means the value of the parameter when $\sfn=S_{\bli,49}$.  To match the numerical $\Sblifn$, we find
\beqa
\hspace{-0.5cm}\phi_{\rm P,eff,bli}&=& 0.06\left(\frac{\xi_{c0}}{0.3}\right) \times \nonumber \\ &&\hspace{-1.5cm}\left\{1+\left[0.4-0.8\left( 1+\frac{M_6}{2}\right)^{-2} \right]\left(1-\log^2\Sigma_2\right) \right\} ,
\label{eq:phi-P-eff-bli}
\eeqa
and $f_{\rm ion,bli}$ is the Draine value (Appendix A) evaluated at $S_{\rm bli,49}$, $M_6$, and $\Sigma_2$. Alternatively, $f_{\rm ion,bli}$ can be approximated by $f_{\rm ion,bli}=0.98 - 0.05M_6$.
 $C_{\bli}$ is given by Equation (\ref{eq:C}) with $\phi_{\rm P,eff}=\phi_{\rm P,eff,bli}$. This analytic fit to $S_{\rm bli,49}$ is 
accurate to better than a factor of 1.5 for the entire parameter space of $\Sigma_2$, $M_6$ and $\xircz$ except for the tiny region $M_6\sim 10$ and $\Sigma_2\sim 0.8-1.0$, where the error increases to $\sim 1.7-1.8$.
Note that  $\phi_{\rm P,eff,bli}\sim 0.05$ ($\xircz/0.3)$ over a large range of the $M_6-\Sigma_2$ parameter space.   This corresponds
to an average cloud pressure, $\bar P_s$, that is equal
to the cloud pressure at $R=0.95R_c$, which
is quite reasonable for shells propagating in the $\theta=0^{\circ}$ direction from $R=
(1-\xircz)R_c=0.7R_c$ (for $\xircz=0.3$) to $R_c$.


Equation (\ref{eq:Sblinew}) shows that usually only very small (low $\sfn$) associations stay embedded  unless $\Sigma$ and $M$ are large.  Utilizing the analytic approximations for $\xireq$ and $\teq$, one can show that nearly all associations in  our $M_6-\Sigma_2$ parameter space that stay embedded come into pressure equilibrium with the surrounding cloud (i.e., $t_{\rm stall}<\tion$). Only for high $M_6$ and low $\Sigma_2$ do associations die before
their shells reach pressure equilibrium
while embedded in the cloud.

\subsubsection{Blister ($\xircz<\xi_{cs,f}<\xircom$, $S_{\rm bli}<S<\Scom$)}

HII regions that expand outside the cloud are termed blister HII regions. Such HII regions can be created in one of two ways. First, a blister HII region will be created if the association is sufficiently luminous that the initial flash of ionization extends beyond the cloud ($\xirsto>\xircz$); according to Equation (\ref{eq:xisto}), this occurs for
\beq
\sfn>\SHII=22.7 \fion^{-1}\left(\frac{\xircz}{0.3}\right)^3 \Sigma_2^{3/2}\Mcs^{1/2}.
\label{eq:flash}
\eeq
The blue line labeled $\Sflash$ in Figure \ref{fig-6} 
is the minimum value of $\sfn$ for associations that begin their evolution in the blister category and are never embedded.
Figure \ref{fig-6} shows that associations with the maximum possible ionizing luminosity in their host cloud, $S=S_{\rm max}$, start their evolution in the blister category for $0.01<M_6<0.4$ and end in the cometary cloud category with  nearly complete destruction of their natal cloud (i.e., the remnant cometary cloud is very small).  However,
such associations only partially destroy more massive clouds. In fact, in a cloud with $M_6>0.4$, an association with $S=S_{\max}$ starts and ends its shell evolution in the blister category.

Second, and more commonly, blisters are created by HII regions initially embedded, but whose pressure is sufficient to drive the shell of neutral gas surrounding the HII region beyond $\xircz$, or  $\Sblifn < \sfn < S_{\rm flash,49}$.   

Having inferred $\phi_{\rm P,eff,bli}$ and
$\Sblifn$, we are in a position to find
an expression for $\phieff$ that gives a good match to the numerical $M_{\rm loss}$ in the blister stage:
\beq
\phi_{\rm P,eff} =\phi_{\rm P,eff,bli} \, \left[\frac{S_{49}}{S_{\bli,49}} \right]^{b_{P}},
\label{eq:phi-P-eff}
\eeq
where
 \beq
  b_{P}= \frac{0.28 \, (\Sigma_2/10)^{-0.4}}{(M_6/0.5)^{0.5} + (M_6/0.5)^{-0.5}}.
  \label{b-P}
   \eeq
Note that $\phieff$ increases with $\sfn$, 
 since higher values of $S$ drive the shell into the inner parts of the cloud, where the cloud pressure is higher. These equations allow one to determine if and where a blister HII region stalls from Equations (\ref{eq:seq}) and (\ref{eq:xieq}). 

When does the blister stage end?
The dividing line between blisters and the next category, cometary HII regions, is somewhat arbitrary; we adopt the criterion that a blister must have $\thecs < 150^\circ$ or, equivalently, $\xircs < \xircom $,  where
\beq
\xircom= (1.61, 1.54, 1.47) \mbox{~~for~~} \xircz=(0.2, 0.3,  0.4)
\label{eq:xircom}
\eeq
respectively.
We set $\Scom$ as the critical value of $S$ dividing these two categories.

There are actually two criteria to reach a boundary between the different stages of evolution of HII regions: First, the ionizing luminosity, $S$, must be large enough to drive the HII region to the boundary and, second, that must occur before $\tion$. We have found that for an HII region to reach the blister stage, the lifetime criterion is relevant only in a small region of parameter space (high-mass clouds with low surface densities, $M_6\ga 10 \Sigma_2^{0.3}$), so we have ignored that. However, in the case of $\Scom$, both criteria are important. In order for the
shell to reach $\xircs=\xircom$, $S$ must be large enough that (1) the shell does not stall first and (2)  it reaches $\xircs=\xircom$ before $\tion$. Since both conditions must be satisfied,  $\Scom$ is the larger of these two criteria.  One can derive analytic equations for these two criteria, but they are
transcendental and not needed for our analytic solution for $M_{\rm loss}$. 
Instead, we give an approximate parametric fit to
the numerical values of $\Scom$, valid for $\xircz=0.2-0.4$:
\begin{equation}
 S_{\rm com,49} = 12\, M_6^{1.7} \left[ 1 + 5 \, \Sigma_2^{2.5} + \frac{50}{1+(M_6/0.1)^{-5}} \right].
 \label{eq:Scomfit}
\end{equation}

Since $S_{\rm com}$ can not exceed $S_{\rm max}=440 \epsilon_{\rm a,-1}M_6$, one can equate Equation (\ref{eq:Scomfit}) to $S_{\rm max}$ to find the boundary of the region in the $M_6-\Sigma_2$ parameter space   in which the blister HII region survives and the cloud never enters the cometary regime. However,
this solution is transcendental, so we make a parametric fit to this boundary
for $\epsilon_{\rm a,-1}=1$.
We call this boundary $M_{\rm survive,6}(\Sigma_2)$.  Cometary clouds exist only for $M_6< M_{\rm survive,6}(\Sigma_2)$ and $\Scom<S<S_{\rm max}$.  For $M_6> M_{\rm survive,6}(\Sigma_2)$ the cloud never reaches the cometary stage and never loses more than about 70\% of the cloud mass (it ``survives"), even with an association with the maximum ionizing luminosity possible for the cloud.
 An approximate parametric fit for the boundary is 
\begin{equation}
M_{\rm survive,6} \simeq \frac{14 \,\Sigma_2^{-3.4}}{1 + 26 \,\Sigma_2^{-3.25}} \, .
\label{eq:Msurvive}
\end{equation}
 For $M_6 < M_{\rm survive,6}$ the parametric fit in Equation (\ref{eq:Scomfit}) agrees with the numerical value of $S_{\rm com,49}$ within a factor 1.25. 
 
For Milky Way type GMCs, with $\Sigma_2\sim 1$, Equation (\ref{eq:Msurvive}) gives $M_{\rm survive,6}\simeq 0.5$; cometary clouds cannot exist above this initial cloud mass, as seen in Figure \ref{fig-6}. (Numerical results, as opposed to the fit in the equation above, give $M_{\rm survive,6}\simeq 0.4$ for Milky Way clouds.) 
If $\Sigma_2\ga 2.8$, 
the above equation shows that $M_{\rm survive}$ rapidly decreases with increasing $\Sigma_2$.

Figure \ref{fig-6} displays both the numerical $S_{\rm com,49}$ and our parametric fit to $S_{\rm com,49}$
for Milky Way clouds. In the range $0.01<M_6<
M_{\rm survive,6}$ the agreement is within a factor 1.15 and the values of $M_{\rm survive,6}$ disagree only by a factor 1.23. Notable  is the slope $\simeq 1.7$ of $\log \sfn$ vs $\log M_6$ at both extremes
$M_6\ll 0.04$ and $M_6\gg 0.1$,  as indicated in Equation (\ref{eq:Scomfit}).  However, in
the intermediate regime $0.04<M_6<0.1$ the slope steepens.   These three regions correspond to the shell stalling at
$M_6\ll 0.04$; the shell dying at $\tion$
but with $S_{\rm com,49}<10$, so that the 
lifetime, $\tion$, decreases with increasing $\Scom$;
and finally the shell dying
but with a constant lifetime,
$\tion= 4$ Myr since $S_{\rm com,49}>10$.

Although $\Scom$ is not needed for our analytic solution for $M_{\rm loss}$,
the time $\tcom$ for the shell to reach the cometary stage ($\thecs=150^\circ$ or $\xircs=\xircom$) is needed.
The HII region enters the cometary stage only if the shell does not stall ($\xi_{\rm stall}>\xircom$), nor does the association die ($\tcom<\tion$). If these conditions are satisfied, then the HII region enters the cometary stage and 
\beq
\tcom=t_s(\xircom)
\label{eq:tcom}
\eeq
from Equation (\ref{eq:t}). See 
Table 1 or discussion in Section 2.1, approximation 9 for $\xircom$.

\subsubsection{Cometary cloud ($\xircom<\xircs$, $\sfn>\Scom$)}

For larger associations with higher values of
$S$, the shell is driven to $\thecs>150^\circ$. 
As discussed previously, the ejected neutral shell
in these directions is considered to be the cometary cloud.  We continue evaporating the cometary cloud, but do not count
the ejected neutral shell in these directions as mass loss from the initial cloud.   Because we consider all initial cloud mass at $\theta<150^\circ$ as mass loss and there is some evaporation at $\theta>150^\circ$, the resultant cometary cloud initially has mass $\sim 0.2-0.3M$ (insensitive to $\xircz$ in the range $\xircz=0.2-0.4$) which dwindles with time as evaporation proceeds.   Numerically, we follow
the cometary cloud as it is driven away
from the association until $t_{{\rm com},f}$,  the minimum
of $\tion$ and $2\tcom$ (Eq. \ref{eq:tcommax}). Analytically, 
we freeze the shell at $\thecs=150^\circ$ and evaporate
the stationary shell until $t_{{\rm com},f}$.   Since the
mass loss rate for a partial spherical shell goes as $r_s^{1/2}$, this slightly underestimates the mass loss from a shell with constant $\thecs$; on the other hand, photoevaporation compresses the shell \citep{bert90}, which reduces $\thecs$ and tends to compensate.

\subsection{Analytic EUV-Induced Mass Loss from GMCs}
\label{sec:ejection}

Appendices D and E provide the details of the analytic approximations and solutions to the mass loss from GMCs.  
The analytic model for mass loss differs in part from the numerical model
in its splitting of the mass loss into components.  Recall that in the numerical treatment, $M_{\rm loss}= M_{\rm evap} + M_{\rm ej}$. The first term is the ionized mass evaporated {\it to the ISM} (it is equivalently
$M_{\rm ion} - M_{\rm ion,os}$, the total HII mass flowing off the shell minus the amount that remains in the cloud). The second term
is the ejected neutral mass as shell passes through initial cloud surface to the ISM.
We find it simpler for the approximate analytical model to divide mass loss into three terms (Eq. \ref{eq:mloss}), $M_\loss=M_\init(<\thecs)+M_\ionz(>\thecs)-M_{\rm ion,os}$. The first term is the total initial cloud mass that lies
at $\theta<\thecs$: $M_\init(<\thecs)=M_{\rm ej} + M_{\rm ion}(<\thecs)$, all the neutral ejected mass (no neutral gas is ejected at $\theta>\thecs$) plus the HII mass flowing off
the shell at $\theta<\thecs$. 
The second term is the HII mass flowing off  at $\theta>\thecs$ so that
$M_{\rm ion}$, used in Section 4, is equal to $M_{\rm ion}(<\thecs) + M_\ionz(>\thecs)$.   The third term
subtracts the part of the HII mass that remains in the cloud, between $A_o$ and $A_s$ and does not escape to the ISM. The two methods of dividing
$M_{\rm loss}$ are equivalent.  Appendix E provides analytic solutions to each of the three mass loss terms.
They are all dependent on $\thecs$  and therefore time dependent. 

Like the above analytic work for the dynamics, the analytic mass-loss model assumes the shell is a partial spherical shape, and
that we can approximate the shell expansion with 
constant values of $\viieff$ and $\phiiieff$.
It uses the parameter $\phieff(S,\Sigma,M,\xircz)$  to model the possible stalling of the shell due to cloud plus ISM pressure. 
It provides a procedure for generating  an analytic solution
for $M_{\rm loss}/M$
over a wide range of cloud parameters
($M$, $\Sigma$) and association parameters ($\xircz$ and $S$),  that give good fits to the numerical results.
We present these analytic and numerical results in the next sections.

\begin{figure*}
\hbox{\includegraphics[width=18 cm,
angle=0, clip] {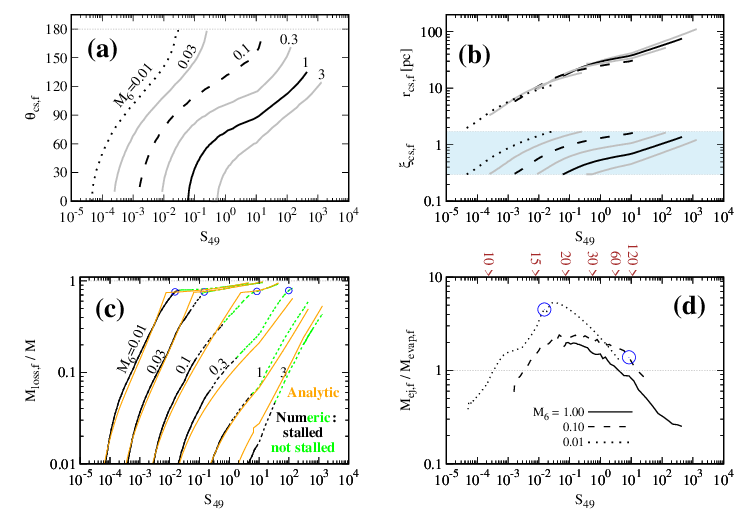}} \caption{
The final state ($t=t_{\rm ion}$, unless cometary clouds with $t_{\rm com,max}$) of shell evolution  around associations  of various luminosities, $S_{49}$, placed at $\xi_{c0}=0.3$ in clouds of various masses, $M_6$. We assume the Milky Way relation $\Sigma_2=1.05M_6^{0.08}$.  Panel a shows  the final  opening angle of the shell, $\theta_{cs,f}$. The dotted, dashed, and solid lines denote the different cloud masses as labeled, and are repeated in panels b and d. Panel b shows the final distance from the association to the shell at the  cloud surface in terms of the normalized radius, $\xi_{cs,f}$, 
(from 0.3 to 1.7, blue area) and the radius in pc, $r_{cs,f}$.  
Panel c shows the total final mass lost, $M_{{\rm loss},f}$, from the cloud by both neutral shell ejection and HII evaporation to the ISM  divided by the initial cloud mass $M$. The orange lines are the analytic model values. The black and green curves correspond to the numerical results. When the curve is black, the shell has stalled in the 180$^o$ direction before $t_{\rm ion}$. When the black curve is solid, the stalled shell gets completely ionized after stalling but before $t_{\rm ion}$. When the curve is green, the shell never stalls in the 180$^o$ direction. The green curve is solid if the shell is completely ionized before $t_{\rm ion}$ while moving.
Panel d shows the ratio of the neutral mass ejected, $M_{{\rm ej},f}$, to the ionized gas evaporated from the cloud into the ISM, $M_{{\rm evap},f}$, for $M_6=0.01$, 0.1 and 1.  For high-mass clouds, evaporation tends to dominate, whereas for low-mass clouds ejection dominates. The open circles in  c and d identify $S_{49}=S_{\rm com}$, or equivalently $\theta_{cs,f}=150^\circ$.  The luminosity of individual massive stars is labeled at the top of panel d.
}
\label{fig:Mloss} 
\end{figure*}

\section{Mass Loss in Milky Way Type GMCs}

Figure \ref{fig:Mloss} provides numerical and analytic results for GMCs in the Milky Way
for associations at $\xircz=0.3$ --i.e., at a radius $R=0.7R_c$. Recall that we assume the Milky Way relation of $R_c$ to
$M$, so that the surface densities corresponding to GMC masses $M_6=1$, 0.1, 0.01 are $\Sigma=105$, 87, and  73  M$_\odot$ pc$^{-2}$.   We terminate
the curves at $S_{\rm max}$, the luminosity of an association with stellar mass equal to $0.1M$, which we take as our upper bound on association
mass and EUV luminosity.  Note that because larger clouds can have larger associations, the curves reach higher $\sfn$ as $M_6$ increases.
Panels a, b and d in the figure present numerical results, while panel c presents both 
numerical and analytic results.

Panel $a$  shows that $\theta_{\rm cs,f}$, the final angle of the ray from the association to the edge of the blister,  first increases rapidly as $\sfn$ increases above $\Sblifn$ and the HII region breaks out of the embedded stage. For low-mass clouds with moderate to high $S$, the final value of
$\theta_{\rm cs,f}$ reaches 180$^\circ$, the opposite side of the cloud, where  $\xircf=1.7$.  Recall that such clouds entered the cometary stage when $\thecs$ reached $150^\circ$. In massive clouds, even with $\sfn=S_{\rm max}$, the shell does not reach the opposite side of the cloud in a time $\tion$.
To reach a given $\theta_{\rm cs,f}$ requires higher $\sfn$ as cloud mass increases, because of their larger size. 

Panel b presents the final shell distance (at $\thecs$) from association   in both parsecs ($r_{\rm cs,f}$) and normalized ($\xi_{\rm cs,f}$) to the cloud radius, $R_c$.  For a given $\sfn$ the physical distance at $\tion$ is insensitive to cloud mass (because cloud ambient density is relatively insensitive to cloud mass), but the normalized distance varies with cloud mass because
$R_c$ increases with mass.

Panel c shows the dependence of the final mass lost from a cloud on the association luminosity, $\sfn$.   Roughly, $M_{\rm loss,f}/M \propto \sfn^b$, where $b\sim 0.5-0.6$ for  $S_{\rm bli,49}\ll\sfn<\Scom$. We estimate in Section 8 below how $b$ varies
with $M$, $\Sigma$ and $\xircz$. We see a break in the slope at $\Scom$, which occurs near  $M_{{\rm loss},f}/M\sim 0.7-0.8$, where the shell has entered the cometary cloud regime.   There is also a small change in the slope at $\sfn=10$ because of the change in the dependence of $\tion$ on $\sfn$ here (see Eq. \ref{eq:tion}).
The orange lines show the results of the analytic approximation, which lies within a factor of 1.5 of the numerical results and is often considerably closer.   For low cloud mass, $M_6\la 0.5$, and for low $\sfn$, shells can stall and get completely evaporated before $\tion$ (solid black lines), as we have noticed in Fig. 4. 
The reason that the whole shell evaporates at low $\sfn$ is that: (i) the shell stalls at a small distance $r_{\rm stall}$ from the association so that it has not swept up as much mass; (ii) the small $r_{\rm stall}$ means high evaporative mass flux from the shell since
$\nii\propto r_{\rm stall}^{-3/2}$; and (iii) low 
$\sfn$ associations live longer
and thus evaporate longer. After the shell evaporates, the ionization front advances directly into the cloud, with the pressure behind the ionization front balancing the ambient cloud pressure, $P_{\rm II}=\bar P_s$.

Note that both numerical and analytical models show that, for a given fractional mass loss
$M_{{\rm loss},f}/M$, more massive clouds require larger associations (higher $\sfn$),   mainly due to the larger size ($R_c$) of more massive clouds.   For a given fraction $M_{{\rm loss},f}/M$,
the required $\sfn$ goes as roughly
$M^{1.5-2.5}$.   The maximum luminosity of an association in a cloud of mass $M$ is proportional to $M$ (Eq. \ref{eq:smax}), so sufficiently massive clouds ($M_6 \ga 0.5$) can never reach the cometary stage.  The final mass loss is correlated with $\theta_{{\rm cs},f}$, since all the mass within the cone defined by this angle for $\theta_{\rm cs,f}<150^\circ$ is lost to the ISM.

\begin{figure*}
 \centering
 \hspace*{-1.0 cm}
  \hbox{\includegraphics[scale=1.2,angle=0]{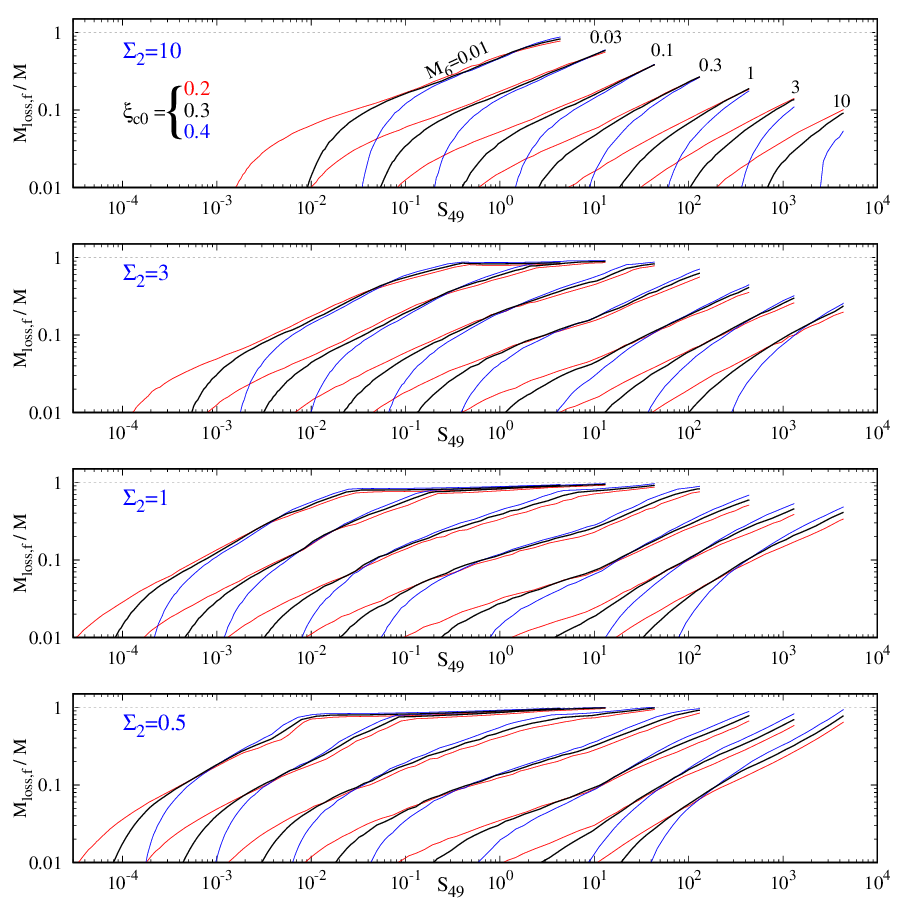}}
\caption{Numerical results for $M_{{\rm loss},f}$ as function of $S_{49}$ for $\xi_{c0}=\{0.2, 0.3, 0.4\}$, $M_6=\{0.01, 0.03, 0.1, 0.3, 1, 3, 10\}$ and $\Sigma_2=\{0.5, 1, 3, 10\}$. The results are shown for $\xi_c0=$0.2 (red), 0.3 (black), and 0.4 (blue). Note for low mass clouds at relatively high $\sfn$ the change in slope indicating cometary cloud phase.  Here, $\sim 0.1-0.2 M$ of the cloud survives as a cometary cloud.
 }
\label{Fig: Mloss-Numerical-Model}
\end{figure*}

  Panel d shows the ratio of 
the neutral shell mass ejected to the HII mass evaporated
from the cloud into the ISM.   Notable is that photoevaporation dominates ejection of neutral gas for higher $\sfn$ associations in massive ($M_6\sim 1$) clouds, but
ejection tends to dominate in the lower mass clouds. The ratio $M_{\rm ej}/M_{\rm evap}$ declines with $\sfn$ at high $\sfn$ for all three cloud masses.  Note for the lower mass clouds that the decline happens once $\sfn>\Scom$, corresponding to $t_{\rm com} < t<\tion$, so that the cloud is in the cometary stage and mass loss is entirely due to photoevaporation in this stage.
At lower $\sfn$ for
$M_6=0.01$ and 0.1, the
loss cone defined by $\theta_{{\rm cs},f}$
increases rapidly with $\sfn$, which enhances the
neutral mass that is ejected relative to
evaporation.  However, the main point of this panel is to note the large range of parameter space ($M$, $\sfn$)
where $M_{{\rm ej},f}/M_{{\rm evap},f} >1$.
If associations in a cloud largely lie
in this domain, ejection of neutral gas dominates the lifetime of the GMC.

\begin{figure*}
\centering
\hbox{\includegraphics[scale=1.2,angle=0]{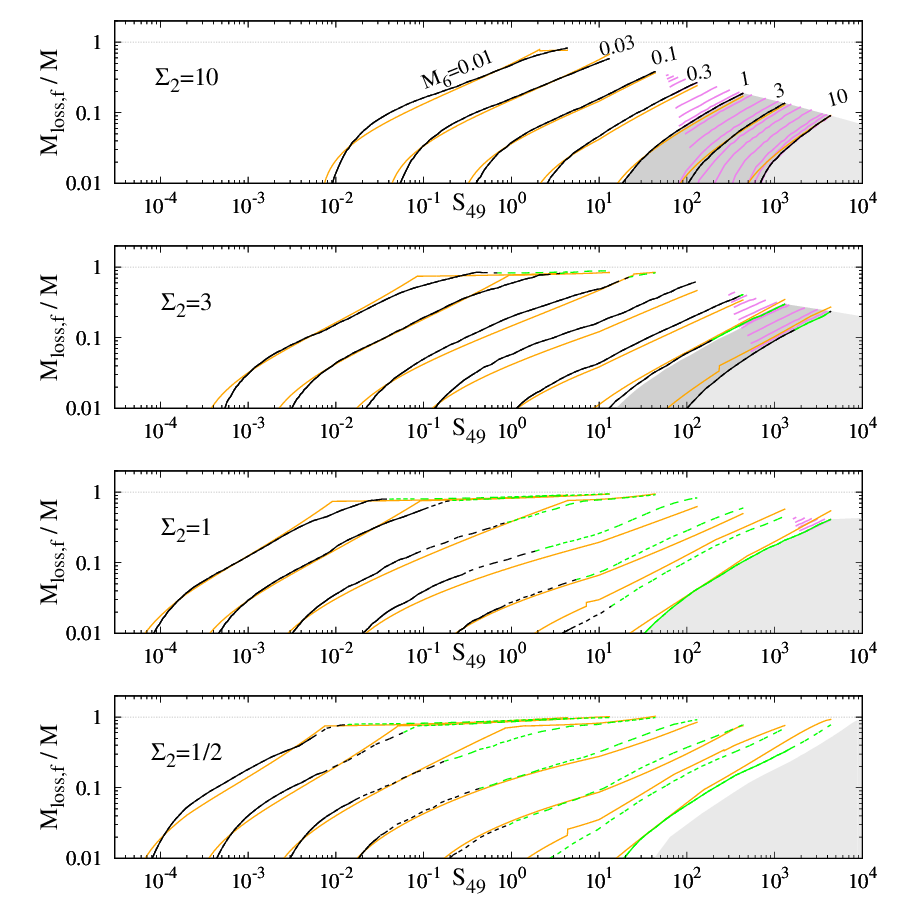}}
\caption{Mass loss in the numerical and analytic models for the case $\xi_{c0}=0.3$. The black and green curves correspond to the numerical results from the model for the evolution of HII regions described in \S 4. When the curve is black, the shell has stalled in the 180$^o$ direction before $t_{\rm ion}$. When the black curve is solid, the stalled shell gets completely ionized after stalling but before $t_{\rm ion}$. 
When the curve is green, the shell never stalls in the 180$^o$ direction. The green curve is solid if the shell is completely ionized before $t_{\rm ion}$ while moving. 
The orange curves correspond to the results generated by the analytical model described in \S 5 and in Appendices D and E.
The gray areas indicate the parameter space for which gravity plays a dominant role in slowing the expansion of the ionized gas;  gravity is not included in our calculations.  The light-gray areas at the right of the $M_6=10$ curve crudely indicate the region where gravity plays a dominant role outside our   parameter space.  The violet hatched areas  indicate the parameter space for which radiation pressure dominates thermal pressure. In the case $\Sigma_2=1/2$ radiation pressure never dominates, and the condition for gravity domination starts at $M_6=20$, outside the parameter space considered here. We have not extended the violet hatched area beyond the range of our calculations  for $M_6>10$, since for $\Sigma_2=0.5-3$ the lower boundary is uncertain.
}
 \label{Fig:Mloss_a}
\end{figure*}

\section{Mass Loss in GMCs with $0.5<\Sigma_2<10$}
\label{sec:Sigma}

In order to extend our results to external galaxies 
(and to GMCs in the Milky Way that differ from the mean found by Rosolowsky et al (in preparation)), we show in Figure \ref{Fig: Mloss-Numerical-Model} the numerical results for mass loss in GMCs over a wide range of cloud parameters, $\Sigma$ and $M$,
and for associations with a wide range of $\sfn$ that are  placed at three positions  in the cloud, $\xircz=0.2$, 0.3 and 0.4.    Figure \ref{Fig: Mloss-Numerical-Model} shows that the different $\xircz$ cases diverge 
for a given $M$ only at low $S$ (less than a few percent of $S_{\max}$). This is
because breakout from the embedded state (which has no mass loss) occurs for lower $\sfn$ if the association is closer to the surface (smaller $\xircz$). Notably, the three cases
give similar values of $M_{{\rm loss},f}/M$
at high $\sfn$, corresponding to $M_{{\rm loss},f}>(0.05-0.2)M$, where the exact placement of the association has only a minor effect on $M_{{\rm loss},f}$.  If high luminosity associations dominate cloud lifetime, this translates to EUV-induced lifetimes of clouds that are independent of $\xircz$.

For high-mass clouds, one sees a slight  break in the slope of $M_{{\rm loss},f}/M$ vs $\sfn$ at
$\sfn=10$ because of the increase in $\tion$ as $\sfn$ decreases below 10.
When the cloud enters the cometary stage, the slope decreases greatly for low-mass clouds because ejection stops as a mass loss process.  Here, $\ga 70\%$ of the cloud has been destroyed, but a small fraction of the initial cloud persists as a cometary cloud, even for 
high $\sfn$.   High-mass clouds, $M_6\ga 1$, are so large that the shells never reach the cometary stage, even at
$S_{\rm max}$.  In general,
the slope of $\log(M_{{\rm loss,f}}/M)$ vs $\log\sfn$ between a few times $\Sbli$ and $\Scom$ is in the range 0.45-0.75,  although the curves are not pure power laws.  The 
results show the full range of possible ionizing luminosities for associations in a cloud of given mass.  Note that low-mass clouds have a much wider range of associations that almost destroy the entire cloud, compared to high-mass clouds.

Figure \ref{Fig:Mloss_a} presents the same numerical results as Figure \ref{Fig: Mloss-Numerical-Model} for the case $\xircz=0.3$, but in addition it presents the results of the analytic model described in Appendix E.  After adjusting one  parameter, $\phieff$ (Eq. \ref{eq:phi-P-eff}),   the agreement in $M_{\rm loss}$ between numerical and analytic models is very  good.   In all of the relevant parameter space, including the cases $\xircz=0.2$ and 0.4 (not shown), the agreement is better than a factor of 1.5, except in a small region of parameter space near $\Sigma_2=0.5$, $M_6=3$, and $\sfn=3$, where
it is better than a factor 2.

In addition,  for the numerical model, Figure \ref{Fig:Mloss_a} shows which cases  have the shell stalling before the association dies at $\tion$, and which cases do not.  For $\Sigma_2\la 3$,  it is the more massive clouds in which the shells do not stall, because the clouds have
lower density and the associations  can have higher luminosities and shorter lifetimes to drive the shells. Shells tend to stall in low-mass clouds since they have smaller associations, and those with $\sfn<10$ live longer, providing more time to reach the stall condition.
The figure also shows it is more difficult for associations to stall in low-$\Sigma$ clouds since the pressure is lower there (Eq. \ref{eq:bPs}). 

Finally, Figure \ref{Fig:Mloss_a} shows the region of high $M$, $\Sigma$ and $S$ where radiation pressure dominates thermal pressure.   Our results do not apply in this regime.  As discussed in Appendix B, we have made a parametric fit to our numerical results for the condition that radiation pressure is dominant over thermal pressure (Eq. \ref{eq:Sch}).
In addition, the figure shows in shaded grey the regions of high $M$ and $\Sigma$ where gravity can impede the evaporation and lower the mass loss rate (see Eq. \ref{eq:grav}).   These two processes
overlap considerably in parameter space and they oppose each other, so that in these overlapping regions it is unclear whether our result is a lower or upper limit.

\section{Parametric Fit to Final Mass Loss}

Although the main text and Appendixes D and E provide  a prescription for  an analytical 
approximation to the numerical results for $\Sbli$ and $M_{\rm loss}/M$ based on physics and 
one physically motivated parameter,
$\phieff$, for the reader's convenience we directly fit the numerical results to produce
a simple equation for $M_{\rm loss,f}/M$.
We call the former our ``physical fit" or ``analytic model" and the
latter our ``parametric fit".   
 Our physical fit is more accurate and produces $M_{\rm loss}/M$ as a function of time, whereas our parametric fit only applies to the final $M_{\rm loss,f}/M$ at $t=\tion$.   Both these fits apply to stalled shells or expanding shells and 
include the variations induced by the cloud
and association parameters $M$, $\Sigma$,
$S$, and $\xircz$.  For $\sfn<S_{\bli,49}$ $M_{\rm loss,f}/M=0$; for $\sfn> S_{\bli,49}$ we give the parametric fit in terms of 
$S_{0.1,49}$,  the value of $\sfn$ that produces $M_{\rm loss}/M=0.1$: 
\beqa
\frac{M_{\loss,f}}{M} &=&0.1 \left(\frac{S_{49}}{S_{0.1,49}}\right)^p  \times\nonumber \\ &&\left\{1-\exp\left[-\frac12\left(\frac{S_{49}}{S_{\bli,49}} -1\right)\right]\right\},
\label{eq:formula-Mloss}
\eeqa
for  $M_{\rm loss,f}/M<0.85$, where $S_{\bli,49}$ is given by Eq. (\ref{eq:Sblinew}) and
\beqa
 S_{0.1,49}&=&\Sigma_2^{5/2} M_6^{3/2} \times\nonumber \\ &&\hspace{-2cm} \left[  0.9\Sigma_2^{-1} + \frac{20 \,\, \Sigma_2^{-1.75}}{ 1 + M_6^{-1.6}  \exp(-0.64 -0.11  \Sigma_2^{1.6})  }\right],  
\eeqa 
\beq
 \hspace{0.25cm} p=(0.5+0.01 \, M_6/\Sigma_2) \, \left(\frac{\xi_{c0}}{0.3}\right)^{0.25}.
 \label{eq:p}
\eeq
This shows that for
$\sfn>S_{0.1,49}$, $M_{{\rm loss},f}/M$ roughly increases as $\sfn^p$, where $p$ ranges from $\sim 0.45$ for $\xircz=0.2$ and a low ratio of $M_6/\Sigma_2$ to 0.75 for $\xircz=0.4$, $M_6=10$, and $\Sigma_2=0.5$. For high values of $\sfn$ that give
$M_{\rm loss,f}/M>0.85$, we set $M_{\loss,f}/M=0.85$, i.e.,
\begin{equation}
    \frac{M_{\rm loss,f}}{M}= \min({\rm Eq}. \ref{eq:formula-Mloss}, 0.85)
\end{equation}
The parametric fit
is good to roughly a factor of 2 in
the broad parameter space of $M$, $\Sigma$, and $\xircz$. It is not as accurate as the physical fit (analytic model),  a factor 1.5,  because it forces a power law behavior when numerically and analytically it is not a pure power law for $M_{\rm loss,f}/M<0.85$. In addition, the capped value of 0.85 (although accurate to a factor of 1.5 for $M_{\rm loss,f}/M$) gives only an approximate value of $0.15M$ for the final masses of the  cometary clouds.  Our numerical and analytical treatments provide somewhat better estimates, but our models in general cannot be expected to give accurate cometary cloud masses, given the model approximations. Finally, the parametric fit does not treat stalled shells and expanding shells separately, like the analytic model does.

\section{Summary}

As discussed in the Introduction, numerous authors including ourselves have concluded that the dominant mechanism destroying GMCs involves the creation of HII regions by the ionizing  luminosity, $S$, from massive stars in associations, and the subsequent evolution of these HII regions.    Photoionization makes the temperature jump by a factor $\sim 100-1000$, which generally causes the HII region to break out of the cloud. Most of the ionized gas is lost to the ISM, a process termed photoevaporation. 
In addition, the expanding HII regions drive neutral shells to escape speed (``ejection"), and they dissipate in the ISM.

In order to model a wide range of GMCs and the OB associations that 
drive mass loss from them, we have 
constructed an approximate, simple
(relative to 2D or  3D hydrodynamical
models) numerical model. In addition,  we present an approximate analytic model that provides good agreement to the more accurate numerical results and provides insights into the parameter dependence of the mass loss. This agreement also serves as a check on the numerical results.  In both models we focus on the results for
$M_{\rm loss}(t)$, the accumulated mass lost at time $t$, which is the sum of the evaporated mass plus the ejected mass (Eq. \ref{eq:mloss2} and Appendix E).  Both models are simple enough that we can explore a wide range of the parameter space:
cloud mass, $M$, cloud surface density,
$\Sigma$, association luminosity, $S$,
and association placement in the cloud,  $\xircz=(R_c-R_a)/R_c$.   In the Milky Way case,
we use the observed typical relation of $\Sigma$ to $M$ (Eq. \ref{eq:Sigma}) to reduce the parameter space to $S$, $M$, and $\xircz$.  In the general case,
at very high values of $M$ and $\Sigma$, our parameter space is constrained because radiation pressure and/or gravity significantly affect the mass loss, and our model is no longer valid (Eqs. \ref{eq:Sch}, \ref{eq:grav} or see Figure 9).

Our numerical and analytic models both assume a spherical GMC (radius $R_c$) of constant density (Fig. 1). The numerical model assumes 
a pressure, $P_s(R)$, that supports the initial cloud  against gravity
(Eq. \ref{eq:p0}), whereas the analytic model
ignores $P_s$ for the dynamics of the expanding shell, but approximates a
$\theta$-independent $\bar P_s$  to compute the criterion for stalling (Eqs. \ref{eq:Ps} and \ref{eq:bPs}).  The association is
characterized by its total ionizing luminosity, $S$, summed over all stars in the association.  The association is placed off center in the spherical cloud at distances
$R=0.6R_c$ ($\xircz=0.4$), $0.7R_c$
($\xircz=0.3)$, and $0.8R_c$ ($\xircz=0.2)$ from the cloud center.
The ionizing luminosity of the association, $S$, is defined as the sum of the ionizing luminosities of the stars in the association, each averaged over the main sequence lifetime of the stars. The lifetime of the association, $\tion$, is defined so that $S\tion$ is the total number of ionizing photons emitted by the association over its lifetime. We approximate
the evolution of $S$ with time as being constant for
$t<\tion$ and zero
thereafter.   Small associations do not fully sample the IMF and are deficient in very massive stars, so
their $\tion$ is longer than that of large
associations (Eq. \ref{eq:tion}).

The expansion of the HII region
and the neutral shell of swept up cloud material around it is treated as purely radial motion.  In the analytic model we assume spherical shells (Eq. \ref{eq:xis}). However, in the numerical treatment, which looks at the dynamics of shell segments at various angles $\theta$ relative to the line from 
the association to the nearest point on the cloud surface, the   shell becomes non-spherical within the cloud due to the cloud pressure gradient (Eq. \ref{eq:dvdt}).  
  In both the
analytic and numerical models, the shell can either stall at $t<\tion$,
when the HII pressure equals the cloud pressure (Eqs. \ref{eq:seq}, \ref{eq:xieq}), or the shell can still be expanding at $t=\tion$.  The
stall condition depends on the parameters $M$, $\Sigma$, and $S$.
We find that the shell stalls for a significant range of this parameter space (high $\Sigma$, low $M$ and low $S$), whereas the association dies
while the shell is still expanding in the opposite range (Fig. \ref{Fig:Mloss_a}).

Small associations ($S<\Sbli$, Eq. \ref{eq:Sblinew}) stay stalled and embedded in the GMC their entire lifetime and no cloud mass is lost.  For $S>\Sbli$, the HII region either instantly breaks out of the cloud ($S>\Sflash$, Eq. \ref{eq:flash}), or the HII region expands and breaks out later; in either case, the result is
``champagne," or blister, flow
($\Sbli<S<\Sflash$).   We show that
$\Sbli$ is roughly proportional
to $\xircz^3 \Sigma^{5/2} M^{3/2}$ (Eq. \ref{eq:Sblinew}).
For Milky Way GMCs with $\Sigma_2\sim 1$ and $\xircz=0.3$, we find $\Sblifn\sim 6\times 10^{-5}$ for $M_6=0.01$, corresponding to a single star with a mass of about 8 \msun, , and $\Sblifn\sim 0.2$ for $M_6=1$, corresponding to a single star with a mass of about 25 \msun\   (Fig. \ref{fig:Mloss}). The maximum values of $S_{\rm max,49}$ for these same clouds are 
4.4 and 440 (Eq. \ref{eq:Sigma}), respectively, so there is a large range of ionizing luminosities that cause mass loss in
these clouds.  
For $S>\Scom$ (Eq. \ref{eq:Scomfit}) the expanding shell at late time enters the cometary cloud regime, which occurs roughly at the time when the accumulated mass loss fraction $M_{{\rm loss}}/M \sim 0.7-0.8$.   For Milky Way type GMCs, $S_{\rm com,49}\sim 10^{-2}$ for $M_6=0.01$ clouds and $S_{\rm com,49}\sim5 $ for $M_6=0.1$ clouds
(Fig \ref{fig:Mloss}).
Even the largest association possible does not drive an $M_6 \ga 0.5$ cloud into the cometary stage in typical Milky Way clouds. 

Massive clouds require larger associations to effect equal fractional mass loss as in lower mass clouds (Figs \ref{Fig: Mloss-Numerical-Model}, \ref{Fig:Mloss_a}).  This is primarily due to their larger size, $R_c$, which requires higher ionizing luminosities to drive a shell to the same value of $\xi_{cs,f}$, the fractional distance from the association.  The scaling is roughly 
$S\propto M^{1.5-2.5}$.  Since
$S_{\rm max}$ is proportional 
to $M$ (Eq. \ref{eq:smax}), a lower power, the above scaling means that low-mass clouds have a large range of possible ionizing luminosities that almost completely destroy the cloud, whereas massive clouds ($M_6>M_{\rm survive,6}$, Eq. \ref{eq:Msurvive}) do not contain sufficiently luminous associations to even get them to the cometary cloud stage. We present as a function of $\Sigma$ a parametric fit to $M_{\rm survive}$ (Eq. \ref{eq:Msurvive}), the critical cloud mass above which cometary clouds cannot form and at least $\sim 30\%$
of the cloud survives even with the most luminous association possible.
A large range of associations ($\Sbli<S<\Scom$) end their evolution in the blister or champagne stage, neither embedded nor
in the cometary cloud stage (Figs \ref{Fig: Mloss-Numerical-Model}, \ref{Fig:Mloss_a}).

Overall, as expected, $M_{{\rm loss},f}/M$ rises with increasing $S$
for $S>\Sbli$.
From $S\sim 3\Sbli$ to the smaller of
$S_{\rm max}$ or $\Scom$, a power law fit to the numerical results for the mass loss gives $M_{{\rm loss},f}/M \propto (S/\Sbli)^p$ with $p\sim 0.45-0.75$, depending 
primarily on $M$ and $\Sigma$.   For
$S>\Scom$ the slope $p$ decreases
appreciably.  Section 8 gives a simple mathematical fit to the dependence of $M_{{\rm loss},f}/M$ on $\sfn$ and the other parameters.

For Milky Way type GMCs, we find that the mass loss in low-mass clouds is largely due to the ejection of neutral shells into the ISM, whereas in high-mass clouds with moderate to high $\sfn$ the photoevaporation dominates the mass loss (Fig \ref{fig:Mloss}d).  Therefore, especially in computing lifetimes of low-mass clouds, it is vital to include the ejection process.

\section*{acknowledgments}
The research of CFM was supported in part by the NASA ATP grant 80NSSC20K0530.  We thank Suzanne McKee for preparation of Figures 1 and \ref{Fig-E}, and the referee for helpful comments.

\bibliographystyle{aasjournal}
\bibliography{destr}{}

\clearpage

\appendix
\section{EUV fraction $\fion$ Absorbed by Gas}
\label{app:fion}

The fraction of EUV radiation absorbed by the gas, $\fion$, varies from unity at low dust optical depth in the HII region to smaller values as the dust optical depth
becomes significant.  The dust
optical depth,  $\tau_{d0}$, in a constant density HII region in which dust absorption
is ignored (hence the 0 in subscript) is \citep{drai11b}
\begin{equation}
   \tau_{d0}= 0.21(\sfn \nii)^{1/3}
   \left(\frac{\sigma_d}{10^{-21}\ {\rm cm^2}}\right),
\end{equation}
where $\sigma_d$ is the average dust cross section in the photon energy range 5 eV-30 eV.  
We assume that H is fully ionized, but He is neutral, so the ion density, $\nii$, is the same as the density of H nuclei, $n_0$.  The dust cross section in HII regions is quite uncertain. 
\citet{drai11b}'s fiducial value, $\sdmto\simeq 1.0$, is somewhat less than the standard value for the diffuse ISM, $\sdmto=1.5$. He also considers values $\sdmto\simeq (0.5,\, 2.0)$. \citet{salg16} provide a measurement of the cross section for the Orion bar that is considerably smaller, $\sdmto\simeq 0.2$, although this is only for the FUV portion of the spectrum; including the ionizing radiation, this would become $\sdmto\simeq 0.3$ based on Draine's results. Here we shall adopt Draine's low value, $\sdmto=0.5$. 

Draine's results depend on two key parameters. The first,
$\beta$, is the ratio of the $h\nu<13.6$ eV luminosity to the $h\nu>13.6$ eV
luminosity.   Like Draine, we adopt $\beta=3$, which corresponds to a 32,000 K blackbody. 
The second parameter, $\gamma$, is proportional to $\sigma_d$, and for our assumed value, $\gamma\simeq 5$.
Given these parameters, \citet{drai11b} finds
\begin{equation}
    \fion\simeq \frac{1}{1+0.84 \tau_{d0}} + \frac{0.18 \tau_{d0}}{1+0.41 \tau_{d0}},
    \label{eq:fion}
\end{equation}
which we adopt in this paper. For the numerical model, we evaluate this at each time step; for the analytic model, we use the value at $t=0$. 
This equation shows that $\fion$
ranges from unity at low dust optical depth to  0.44 at high optical depth.   We note that
 WM97 assumed a constant $\fion=0.73$ as an average over the Galaxy of all HII regions.

 \section{Radiation Pressure}
 \label{app:rp}
\setcounter{equation}{0}
 
 In the text, we focus on gas pressure as the dominant force driving the expansion of HII regions. As shown by  \citet{krum09} and updated by \citet{lope11} and \citet{jeff21b}, direct radiation pressure -- i.e., the pressure due to stellar radiation -- can also be important, although primarily on relatively small scales.  In a study of almost 5000 HII regions in M83, \citet{dell22} found that direct radiation pressure was almost never important.  
(Indirect radiation pressure -- i.e., that due to reprocessed stellar radiation in the IR -- is important only for a top-heavy IMF or high dust 
abundance--\citealp{skin15}.)
 Under the assumption that radiation pressure has not altered the density distribution in the HII region,  the radius of an HII region at which gas pressure equals radiation pressure at the Str\"omgren radius is 
\beq
\rch=\frac{\ab(L/c)^2}{12\pi(2.1\phiii kT)^2(f_\ionz S)}.
 \label{eq:rch1}
 \eeq
 The ratio of gas pressure to rad pressure varies with $r$ and radiation pressure is dominant at small $r<r_{\rm ch}$.
 We have altered the result of \citet{lope11} by including the factor $\phiii$ for the ram pressure due to photoevaporation and omitting the trapping factor, which is not relevant for the large HII regions we are considering.
 We also changed the factor 2.2 in their expression, which is the number of particles per H nucleus, to 2.1 since we are assuming that the He is neutral. 
  Following \citet{krum09},
 we define $\psi=L/(S\epsilon_0)$, where $\epsilon_0=13.6$~eV is the ionization potential of hydrogen. We then have
 \beq
 \rch=1.40\times 10^{-3}\left(\frac{\psi^2}{\phiii^2 f_\ionz}\right)\sfn~~~\mbox{pc}.
 \eeq

For an association large enough that the IMF is well sampled ($M_a\gg 10^3$ \msun) and for
 the \citet{weid06} main-sequence IMF extending from 0.08~\msun\ to 120~\msun, Starburst99 \citep{leit99}\footnote{We use Starburst 99 in this section since \citet{parr03} did not give values for the bolometric luminosity.} gives $3.4<\psi<10.2$ for 0.01 Myr~$<t<4$~Myr, with an average value over that time interval of $\bar\psi\simeq 5.7$. (\citealp{krum09} took $\psi=1$.) Relative to the cloud radius (Eq. \ref{eq:rc}), this implies
 \beq
 \xi_{\rm ch}=0.35\left(\frac{\eamo\Sigma_2^{1/2} M_6^{1/2}}{\phiii^2f_\ionz}\right)\frac{S}{S_{\max}}.
 \label{eq:xich}
 \eeq

  This result is based on the assumption that radiation pressure is negligible, so that the density in the HII region is uniform. \citet{drai11b} has worked out the effects of radiation pressure on HII regions. We adopt the criterion that radiation pressure can be ignored if the density at the outer edge of the HII region, where it is a maximum, is less than 1.4 times the rms density. For our adopted dust cross section, $\sigma_d=5\times 10^{-22}$~cm$^2$,  one can show that his results imply that radiation pressure can be ignored for $\xi_s>\xi_{\rm crit}$ with $\xi_{\rm crit}\simeq \xi_{\rm ch}$. This is self-consistent: the condition for radiation pressure to be neglected that is derived for an HII region without radiation pressure is about the same as that for an HII region in which radiation pressure is present, but weak. 
 
 To set a parameter boundary on
 where radiation pressure dominates,
 we use the numerical result for
 $\xi_{ s,f}(\theta=180^\circ)$, the final ($t=\tion$ or $2t_{\rm com}$ if $2t_{\rm com}<\tion$) value of $\xi_s$ at $\theta=180^\circ$.   If 
 $\xi_{s,f}(\theta=180^\circ)<\xi_{\rm ch}$, radiation pressure dominates the thermal pressure driving the shell.  A parametric fit to the boundary is given in the main text (Eq. \ref{eq:Sch}).
 In fact, even if $\xi_{s,f}(\theta=180^\circ)=\xi_{\rm ch}$,
  the results of
\citet{krum09} show that
the radius of the HII region is only about a factor 1.3 times larger than the value
without radiation pressure.

\section{Pressure in ISM, $P_{\rm ISM}$}
 \label{app:pism}
\setcounter{equation}{0}

The pressure in the ISM is due to the gas, the magnetic field and cosmic rays. We assume that interchange instabilities substantially reduce the magnetic pressure gradient across the shell and that cosmic rays diffuse so that they do not exert a significant force. As a result, the external pressure acting on the shell of the HII region is due primarily to the pressure of the interstellar gas. We estimate the thermal pressure of the interstellar gas from a fit to the average thermal pressure in the Galactic disk found by \citet{wolfire03}, normalized to agree with the value $5.2\times 10^{-13}$ dyne cm\ee\ observed in the solar neighborhood by \citet{jenk11},
\beq
P_{\rm th}=2.40\times 10^{-12} e^{-R_{\rm gal}/5.4\ {\rm kpc}}~~~\mbox{dyne cm\ee}.
\eeq
Here we have adjusted the radial scale length to correspond to a distance to the Galactic Center of 8.25 kpc instead of the 8.5 kpc assumed by \citet{wolfire03}.  Note that this is the thermal pressure required to maintain the HI in a two-phase medium. 

The gas pressure also includes turbulent pressure. We use the midplane densities of the various components of the ISM from \citet{mcke15} and the HI velocity dispersions measured by \citet{heil03}, 7.1 km s\e\ for the CNM and 11.4 km s\e\ for the WNM. For the H$_2$, we estimate a velocity dispersion of 5 km s\e\ as that required to produce a Gaussian scale height of 74 pc \citep{dame87} for gas embedded in a medium of total density (including stars and dark matter) of 0.10 \msun\ pc\eee\ \citep{mcke15}. Altogether, this leads to a turbulent pressure in the local ISM of $1.7\times 10^{-12}$~dyne~cm\ee. 
By comparison, \citet{boul90} found a local turbulent pressure of $(1.0-1.5)\times 10^{-12}$~dyne~cm\ee. The fact that their estimate is lower than ours is to be expected since they assumed that over half of the local interstellar gas is molecular with a velocity dispersion of 5 km s\e. 
Our results imply that the ratio of total gas pressure to thermal pressure is about 4.3 in the local ISM, and we assume that this is true  throughout the Galactic disk. 
As a result, the total gas pressure in the Galactic midplane is
\beq
P_\ism\simeq 1.0\times 10^{-11}e^{-R_{\rm gal}/5.4\ {\rm kpc} }~~~\mbox{dyne cm\ee}.
\label{eq:pism}
\eeq
At our fiducial radius, $R_{\rm gal}=5.3$~kpc, this gives $P_\ism\simeq 3.7\times 10^{-12}$~dyne~cm\ee. At the solar circle, this $P_\ism=2.2\times 10^{-12}$~dyne~cm\ee,
corresponding to $P_\ism/\kb=1.6\times 10^4$~cm\eee\ K.

 \section{$\dot M_\ionz$ and $\vii$}
 \label{app:miond}
\setcounter{equation}{0}

The rate at which gas is ionized, $\dot M_\ionz$, is proportional to the velocity at which the gas flows out of the IF, $\vii$.
To determine $\dot M_\ionz$ and $\vii$, we express the rate of change of the mass of the HII region in two different ways,
first by considering the rate of change of the volume of the HII region and second by considering the mass flow into
and out of the HII region.
We define the mass of the partially enclosed HII region, $M_{\rm ion,os}$, to be the mass in the region between $A_o$ and $A_s$ in Figure \ref{fig-3}. 
The mass  of the partially enclosed HII region is the sum
of the HII mass at $\theta>\thecs$ and the cone at $\theta<\thecs$ extending from the association to $A_o$. 
The mass of ionized gas at $\theta>\thecs$ and at time $t$ is\footnote{Note that equation \ref{eq:mhii1} (the instantaneous mass) differs from 
$M_{\rm ion}(\theta>\thecs)$ which is the accumulated ion mass generated at $\theta>\thecs$ from $t=0$ to $t$. Some of the latter expands to $\theta<\thecs$  and also beyond $A_o$.}
\beqa
M_\hii(\theta>\thecs)=\int_{\thecs}^\pi d\Omega\int_0^{\rsth}\rhoii(\theta)r^2dr\\=\frac{2\pi}{3}\int_{-1}^\murcs d\mu_r \rhoii(\theta)\rsth^3,
\label{eq:mhii1}
\eeqa
where $\mu_r=\cos \theta$ and we have assumed that the density is given by the Str\"omgren condition, $\rhoiit=\rho_0[\rsto/\rsth]^{3/2}$, and is independent of $r$. 

To evaluate the rate of change of this mass, we need the rate of change of $\rsth$ at constant $\theta$,
\beq
r_s'(\theta) =\ppbyp{\rsth}{t}\bigg{|}_\theta.
\eeq
This is a phase velocity, not a particle velocity. It is related to the velocity of the shell, $v_s$,
which is the same as the shock velocity and is normal to the surface of the shell, by 
$v_s(\theta)= r_s'(\theta)\cos\alpha$, where 
$\cos\alpha=\hat\vecn\cdot\hat\vecr$
and $\hat\vecn$ is the unit normal to the shell surface. 
We now make the quasi-spherical approximation, in which we retain terms of order $\alpha$ but drop terms of order $\alpha^2$;
as a result, we set $\cos\alpha\simeq 1$ and $r_s'(\theta)\simeq v_s(\theta)$. One can show that non-spherical effects lead to changes in
$\rsth$ that are first order in $\alpha$, and we retain those.
The Str\"omgren condition then implies that the rate of change of the density at constant $\theta$ is 
\beq
\ppbyp{\rhoiit}{t}\bigg{|}_\theta\simeq \dot\rho_{\rm II}(\theta)=-\frac{3\rhoiit v_s(\theta)}{2\rsth}.
\label{eq:rhoiip}
\eeq

With this in hand, we can evaluate the rate of change of the mass in the HII region at $\theta>\thecs$,
\beq
\begin{split}
\dot M_\hii(\theta>\thecs)&=\frac{2\pi}{3}\,\murcsd\rhoii(\thecs)\rcs^3 +\int   \rhoiit v_s(\theta) \rsth^2 d\Omega
+\frac 13 \int \dot\rhoii(\theta) \rsth^3 d\Omega,
\end{split}
\label{eq:mcavd}
\eeq
where $\murcs=\cos\theta_{cs}$ is unity when the HII region breaks out of the cloud and -1 when the shell reaches the opposite end of the cloud. 
With the aid of Equation (\ref{eq:rhoiip}), we find that the third term in this equation is 
\beq
-\frac 12\int  \rhoiit v_s(\theta)\rsth^2 d\Omega,
\eeq
which cancels half the second term.
Since the area of the shell in the quasi-spherical approximation is
$A_s=\int \rsth^2\,d\Omega$, Equation (\ref{eq:mcavd})  becomes
\beq
\dot M_\hii(\theta>\thecs)=\frac{2\pi}{3}\,\murcsd\rhoii(\thecs)\rcs^3+\frac12 \avg{\rhoiit v_s(\theta)}A_s,
\label{eq:mdhii}
\eeq
 where $\avg{...}$ is an average over the surface of the shell.

To evaluate the mass in the cone ($\theta<\thecs$), we assume that the density inside the cone is the same as at its surface, 
which is consistent with the results of \citet{york89}. As a result, we have
\beq
M_\cone=\frac{1}{3}\,\rhoiitcs \rcs\murcs A_o,
\label{eq:mcone}
\eeq
where $A_o=\pi \rcs^2(1-\murcs^2)$ is the area of the base of the cone--i.e., of the opening from the HII region to the ambient medium.
Note that the mass of the cone is negative for $\thecs>90^\circ$, thereby canceling the mass that lies outside of $A_o$.
The mass of the partially enclosed HII region between $A_o$ and $A_s$ is
\beq
M_{\rm ion,os}=M_\hii(\theta>\thecs)+M_\cone
\label{eq:mhii}
\eeq
so that evaluation of the time derivatives gives
\beq
\begin{split}
\dot M_{\rm ion,os}\hspace{-0.1cm}&=\hspace{-0.1cm}\frac 12 \avg{\rhoiit  v_s(\theta)}A_s +\rhoiitcs A_o\left(\rcs\murcsd+\frac 12 \dot r_{cs} \murcs\right).
\end{split}
\label{eq:mhiid1}
\eeq

Alternatively, the mass of the partially enclosed HII region grows at a rate $\dot M_\ionz$ due to photoevaporation but declines 
at a rate
$\dot M_{\rm evap}$
due to mass loss out of $A_o$ to the ambient medium,
\beq
\dot M_{\rm ion,os}=\dot M_\ionz -\dot M_{\rm evap}.
\label{eq:mhiid2}
\eeq
Analysis of the results of \citet{york89} shows that the ionized gas emerges from the opening in the cloud  (area $A_o\simeq \pi (r_{cs} \sin \theta_{cs})^2$, see Fig. \ref{fig-3}) with a typical speed $\cii$ {\it relative to the
ambient cloud}. However, the opening itself is moving in the $\theta=180^\circ$ direction
(to the right in Fig. \ref{fig-3}) with speed
\begin{equation}
    v_o= \frac{d(\rcs \murcs)}{dt}.
    \label{eq:vo}
\end{equation}
Note that $v_o$ is negative since the opening
moves in the 180$^\circ$ direction.  
We assume that
the average density of the gas passing through the opening $A_o$ is approximated by
$\rhoii (\theta_{\rm cs})$, the density between the association and
the edge of the opening.   The 
hydrodynamical models of Yorke et al
(1989) support this assumption. This opening defines the leftmost boundary of the partially enclosed HII region in Figure \ref{fig-3}. 
The mass
loss rate out of the opening is then
\beq
{\dot M}_{\rm evap}\simeq \rhoii(\thecs) (\cii-v_o) A_o.
\label{eq:mlossiond}
\eeq
Equating the two expressions for $\dot M_\hii$,
equations (\ref{eq:mhiid1}) and (\ref{eq:mhiid2}), gives
\beq
\dot M_\ionz=\frac 12 \avg{\rhoiit v_s(\theta)}A_s +\rhoiitcs A_o \left(\cii -\frac 12 \dot r_{cs}\murcs\right).
\label{eq:miond}
\eeq

In order to solve the equation of motion, we need a local relation for the ionized mass loss from the shell. Let $\vii(\theta)$ be
the flow velocity of the ionized gas from the shell, so that (Eq. \ref{eq:mionomd})
\beq
\dot M_\ionz=\avg{\rhoiit\vii(\theta)}A_s.
\label{eq:miond2}
\eeq
We now make the approximation that the mass flux through the IF is independent of $\theta$,  so that $\avg{\rhoiit\vii(\theta)}=\rhoiit \vii(\theta)$. This gives our final expression for the IF outflow velocity:
\beq
\vii(\theta)=\frac{\avg{\rhoiit v_s(\theta)}}{2\rhoiit}+\frac{\rhoiics A_o}{\rhoiit A_s}\left(\cii-\frac 12 \dot r_{cs}\murcs\right).
\label{eq:viitheta}
\eeq

 \section{Approximate Analytic Solutions and Fits}
\label{app:analytic}
\setcounter{equation}{0}

In Section 4 and Appendix \ref{app:miond} we outlined the steps in the {\it numerical} code to obtain the mass loss $M_{\rm loss}(t)$ from
the GMC (Eq. \ref{eq:mloss2}). The numerical solution follows the growth of a {\it non-spherical} shell at $\theta>\thecs$ as time progresses, and
includes the variation with time and angle of the relevant parameters $\phiii$, $\vii$, and $\xircs$ (and thus $v_s$ and $\mu_{\rm r,cs})$. The numerical solution also follows the non-spherical shape of $A_s$, and more accurately tracks the effect of dust absorbing EUV.

Here in Appendix \ref{app:analytic},
as described in Section 5, our primary goal is to obtain an {\it analytic} solution to $M_{\rm loss}(t)$ by approximating various parameters
and assuming {\it spherical} symmetry.  $M_{\rm loss}(t)$ depends on $\vii$ which is time dependent and grows as the shell expands. We take $\viieff$ as the mean value of $\vii$ over the time interval considered.  This constant value enables analytic solution to $M_{\rm loss}(t)$.    
We similarly take a mean value of 
of $\phiii$, or  $\phiiieff=2^{1/2}$.
Finally, to approximate the shell stalling in the cloud, we adjust $\phieff$ to give the best fit of the analytic solution for $M_{\rm loss,f}$ to the numerical $M_{\rm loss,f}$. 

As described in Sections 4.4 and 5.2 there are two regimes of mass loss: (i) the blister  regime, where
$\thecs<150^\circ$ (equivalently $\xircs<\xircom$, Eq. \ref{eq:xircom}) and (ii) the cometary cloud regime, where  $\thecs>150^\circ$ (equivalently $\xircs>\xircom$).   If $\sfn<\Scom$ (Eq. \ref{eq:Scomfit}), the entire evolution of the
shell lies in the blister regime.   If $\sfn>\Scom$ the shell evolves in the blister regime until it passes
into the cometary regime at $t=t_{\rm com}<\tion$.  

In our analytic approximation  for the cometary regime, we assume the shell
remains fixed at $\xircs=\xircom$ with $\thecs=150^\circ$  from $t=t_{\rm com}$ to
$\tcomf={\min}(2t_{\rm com}, \tion$).  
In fact, 
the dynamics after $t>t_{\rm com}$ are complicated since the shell accelerates rapidly once it leaves
the cloud due to
the rocket effect, and this happens at $\theta=150^\circ$  before it happens at larger angles.  The net result is that the shell becomes very non-spherical and $r_s$ in the 150$^\circ$ direction is greater than $r_s$ 
in, say, the 180$^\circ$ direction (see discussion in section 5.2.3). 

\subsection{Blister  regime ($\thecs<150^{\circ}$, $\xircs<\xircom$) }

We start by assuming that the HII region has broken out of the cloud ($\rcs>r_{c0}$) so that mass is being lost.  Recall that $\vecR$ is the radius vector from the initial center of the cloud and 
$\vecR_a$ is the location of the association.  
The radius vector from the association is $\vecr=\vecR-\vecR_a$.  The angle between $\hat\vecr$ and $\hat\vecR_a$ is $\theta$, 
and that between $\hat\vecR$ and $\hat\vecR_a$ is $\Theta$; these are related by $r\sin\theta=R\sin\Theta$. Let $z=r\cos\theta$  be the distance measured
from the association along the axis extending from the cloud center through the association.
We define the internal, or partially embedded, HII region as the ionized gas at $z<\rcs\cos\thecs\equiv r_{cs}\murcs$ (i.e., the gas between $A_o$ and $A_s$ in Fig. \ref{fig-3}). 
The mass of the partially embedded (truncated spherical)  HII region is the
density in the gas, $\rho_0(\xirsto/\xircs)^{3/2}$, times the volume,
$(\pi/3) r_s^3(2+3\murcs-\murcs^3)$, or 
\beq
M_{\rm ion,os}=\frac 13 \pi\rho_0 R_c^3(\xirsto \xircs)^{3/2}
(2+3\murcs-\murcs^3).
\label{eq:miios}
\eeq
Recall that the mass of ionized gas lost from the cloud, $M_{\rm evap}$, is $M_{\rm ion}$-$M_{\rm ion,os}$, where $M_{\rm ion}$ is the total
mass of ionized gas.

The total mass that has been lost from the cloud at any time $t$, $M_\loss$, is then the sum of the mass initially in the mass-loss cone defined by $\thecs(t)$, plus the mass of ionized gas
that originates at $\theta>\thecs$, minus the mass in the partially embedded HII region:\footnote{$M_\loss$, $M_\ionz$ and $M_{\rm ion,os}$ are intrinsically time dependent, as is $\thecs$, but we suppress this notation for simplicity.}
\beq
M_\loss=M_\init(<\thecs)+M_\ionz(>\thecs)-M_{\rm ion,os}.
\label{eq:mloss}
\eeq
The first two terms are evaluated below.

After $t_\ionz$, the gas in the partially embedded HII region recombines and cools. We assume it rejoins the cloud and is not part of the mass loss. We neglect the
small amount of gas in the mass loss cone $\theta<\theta_{cs}$ that will be ejected by a supernova at $t\sim t_{\rm ion}$.

\subsubsection{$M_\init(<\thecs)$}

The gas initially in the mass-loss cone ($\theta<\thecs$) is partly ejected neutral shell and
partly ionized gas, either from the initial HII region or 
photoevaporation of  that part of the shell.
Recall that $\theta$ is an angle centered at the association whereas $\Theta$ is centered at the center of the cloud.
The mass initially inside the mass-loss cone--i.e., the mass initially inside an angle $\thecs$--is 
the sum of the mass at $z>\zcs$, which is $\frac 13 \pi \rho_0 R_c^3(2-3\murcs+\murcs^3)$, plus the mass of the cone extending from the association to $\zcs$, which is $\frac 13\pi \rho_0 (\rcs\sin\thecs)^2\rcs\cos\thecs$. Noting that $\rcs\sin\thecs=R_c\sin\Thecs$, that $R_c\cos\Thecs=\rcs\cos\thecs+R_a$, and that $R_a=(1-\xircz)R_c$, this gives
\beq
M_\init(<\thecs)=\frac 13 \pi\rho_0 R_c^3(1-\murcs)\left[1-\murcs+(1+\murcs)\xircz\right].
\eeq
We can re-express this in terms of $\xircs=r_{cs}/R_c$  by
using  Equation (\ref{eq:cos}) in the text and noting that
$\frac 13\pi \rho_0 R_c^3= \frac 14 M$: 
\beq
M_\init(<\thecs)=\frac{M(\xircs^2-\xircz^2)}{4(1-\xircz)}\left\{1-\frac 14\left[(2-\xircz)^2-\xircs^2\right]\right\}.
\label{eq:mej1}
\eeq
This smoothly increases from $M_\init(<\thecs)=0$ at $\xircs=\xircz$, corresponding to $\thecs=0$, to $M_\init(<\thecs)=M=4\pi\rho_0R_c^3/3$ at 
$\xircs=2-\xircz$, corresponding to $\thecs=\pi$. 
The growth of $M_\init(<\thecs)$ with time is obtained by inserting the time
dependence of $\xircs$ into this equation.

\subsubsection{$M_\ionz(>\thecs)$}

Next consider the gas outside the mass-loss cone. The velocity of the gas flowing from the IF adjusts so that the density is governed by the Str\"omgren condition, $\rhoii=(\xirsto/\xi_s)^{3/2}\rho_0$. 
It follows that in the embedded stage, the specific mass of ionized gas is
\beq\begin{split}
M_{\emb,\Omega}=\frac 13 \rhoii (\xirs R_c)^3=\frac{M}{4\pi}&(\xirsto \xirs)^{3/2}\ \ \ \ \ \ (\xircz\geq \xi_s\geq\xirsto).
\end{split}
\eeq

After the HII region enters the blister stage, the evolution of the specific mass of ionized gas
is governed by Equation (\ref{eq:mionomd}) in the text,
\beq
\frac{d \mionom}{dt}=r_s^2\rhoii\vii.
\label{eq:mevom}
\eeq
The value of $\vii$ is derived  in Appendix \ref{app:miond} and given in Equation (\ref{eq:viitheta2}) for the non-spherical case. For analytic work, we assume that the shock front is spherical.  As a result, $\dot r_{cs}=v_s$ and the density of the ionized gas,  $\rhoii$, is constant with angle for $\theta>\thecs$.   
Since $A_o/A_s= \frac 12(1-\murcs)$ for a spherical shell, Equation (\ref{eq:viitheta2}) becomes 
\beq
\vii\simeq \frac12 v_s+\frac12 (1-\murcs)\left(\cii-\frac 12 v_s\murcs\right).
\label{eq:vii}
\eeq
For an embedded HII region ($A_o=0$, $\murcs=1$), this gives $\vii=\frac12 v_s$, as derived in the text. 
For simplicity, in integrating
Equation (\ref{eq:mevom}) to obtain the accumulated ionized specific mass up to the point where the shell has reached $\xirs=\xircs$, we set $\vii$ equal to a constant, 
\beq
\viieff=\left[\frac 12 v_s(\xircz) \vii(\xircs)\right]^{1/2},
\label{eq:viieff}
\eeq
where $\frac 12 v_s(\xircz)$ (see Eq. \ref{eq:vs})  is the value of $\vii$ at the beginning of the blister stage when $\thecs=0$, and where $\vii(\xircs)$ is the value of $\vii$ when the dimensionless radius of the shell is $\xircs$ (Eq. \ref{eq:vii}). This equation is for the common case $S<S_{\rm flash}$; the case of very luminous associations, $S>S_{\rm flash}$, is discussed below.
As in the text, we cap $\viieff$ at $\cii$.

First, consider the case in which
the HII region has not stalled, so that the radius and velocity of the shell are given by equations (\ref{eq:rs}) and (\ref{eq:vs}) in the text, respectively.
We then find that the specific mass of gas photoionized during the blister stage (note this does not include the initial HII region mass, hence the $\Delta$)
outside the mass-loss cone (i.e., for $\theta>\thecs$) is
\beq
\begin{split}
\Delta M_{\rm bli,\Omega}&(\xircs;\xircz)=\left[\frac{4}{3\phiiieff}\right]^{1/2}\frac{\viieff}{\cii}\left(\frac{\xirsto}{\xircs}\right)^{3/4}\left[1-\left(\frac{\xircz}{\xircs}\right)^{9/4}\right]M_\Omega(\xircs)~~~~
 (\xircs>\xircz),
\end{split}
\label{eq:mionbliom}
\eeq
 where $M_\Omega(\xircs)=\frac13 \rho_0 \Rc^3 \xircs^3$.
Note that 
this result is valid only inside the cloud ($\theta>\thecs$).
Because our spherical analytic model has $\xi_s=\xircs$ for $\theta>\theta_{cs}$, the mass of
gas that has been ionized outside the mass-loss cone is
\beq
M_\ionz(>\thecs)\hspace{-0.1cm}=\hspace{-0.1cm}2\pi(1+\murcs)\left[M_{\emb,\Omega}+\Delta M_{\bli,\Omega}(\xircs;\xircz)\right].
\label{eq:miong}
\eeq
Recall that the mass of gas in the initial Str\"omgren sphere is included in $M_{\emb,\Omega}$. 
Furthermore, $M_\ionz(>\thecs)$ is the accumulated mass (from $t=0$ to $t$) that has been {\it generated} at $\theta>\thecs$; this gas expands away from the ionization front, and at time $t$ only some of it remains at $\theta>\thecs$.

Next consider the case in which the HII region stalls at $\xi_{\rm stall}$ at a time $t_{\rm stall}=t_s(\xi_{\rm stall})$ (Eqs. \ref{eq:t} and \ref{eq:xieq}).  In that case, the specific mass ionized prior to
$t_{\rm stall}$ is given by Equation (\ref{eq:mionbliom}) evaluated at $\xircs=\xi_{\rm stall}$. Including the mass of ionized gas produced after the HII region stalls,
the total specific mass of ionized gas for a stalled blister is 
\beq
\begin{split}
 M_{\rm ion,\Omega}(\xi_{\rm stall},t;\xircz)=M_{\emb,\Omega}+\Delta M_{\bli,\Omega}(\xi_{\rm stall};\xircz)+R_c^2\xireq^2\rhoii \viieq (t-t_{\rm stall})~~~~~\mbox{(stalled)},
 \label{eq:stall}
 \end{split}
 \eeq
 where $\viieq$ is $\vii$ evaluated at $t_{\rm stall}$ (Eq. \ref{eq:vii}). The total mass of ionized gas generated outside the mass-loss cone, $M_\ionz(>\thecs$), is obtained from this
 equation by multiplying by $2\pi(1+\murcs)$, as in the case of Equation (\ref{eq:miong}).  Using this value of $M_\ionz(>\thecs$), 
 $M_{\rm loss}(t)$ for stalled shells is given by
 Eq. \ref{eq:mloss}.

What happens if the cluster is so luminous that the HII region begins in the blister stage ($S>S_\flash$, Eq. \ref{eq:flash})?
In this case, $\xircs$ begins at $\xirsto$ and not $\xircz$. 
 HII regions almost never stall when $S>S_\flash$, so we consider the case in which the blister continues expanding
until the cluster dies.
The specific mass of ionized gas is then given by Equation (\ref{eq:mionbliom}) with
$\xircz$ replaced by $\xirsto$ and $\viieff$ evaluated with the initial value of $\vii=\frac12 v_s(\rsto$) instead of $\frac12 v_s(\rcz)$, since the D-type ionization front first forms at $\xirsto>\xircz$. \\

\subsection{Cometary Regime ($\thecs>150^{\circ}$, $\xircs>\xircom$, $\tcom<\tion$ and $t_{\rm stall}$)}

Reaching the cometary regime requires $\sfn>\Scom$  (Eq. \ref{eq:Scomfit}),  or, equivalently,  $\tcom<\tion$ and $t_{\rm stall}$.  
We denote the mass passing through the IF in the cometary regime by $\Delta M_{\rm com}(t)$.   Recall that in our approximation, we terminate evaporation at $\tcomf={\min}(2\tcom,\,\tion)$.
Since the shell remains at $\thecs=150^{\circ}$ and $\xircs=\xircom$ for that time interval,  we have for $\tcom<t<\tcomf$
\begin{equation}
  \Delta M_{\rm com}(t)=2\pi[1+\cos(150^\circ)] R_c^2\xircs^2\rhoii \cii (t-\tcom).
  \label{eq:mioncom}
\end{equation}
Here, $\xircs=\xircom$ and we have approximated $\vii\simeq\cii$ for this large value of $\xircs$.
The total mass loss includes the mass lost before $\tcom$, $M_{\rm loss}(\tcom)$, which is given by Equation (\ref{eq:mloss}) with $\xircs=\xircom$ and with the term $M_\ionz(>\thecs)$ given by Equation (\ref{eq:miong}):
\begin{equation}
    M_{\rm loss}(t)= M_{\rm loss}(\tcom) + \Delta M_{\rm com}(t).
    \label{eq:mlosscom}
\end{equation}
For sufficiently large ionizing luminosities, this expression can exceed the initial cloud mass, and in that case we set $M_{\rm loss}=M$.

\begin{figure*}

\hbox{\includegraphics[scale=0.65,angle=0]{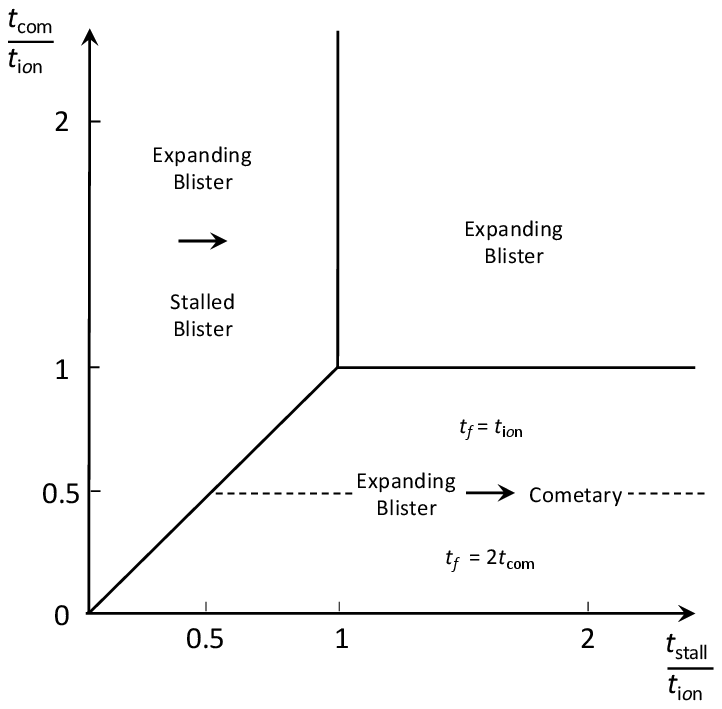}}
\vspace{-12em}
\caption{
Parameter space for HII regions that have broken out of their natal cloud ($r_s > r_{c0}$;  $S>S_{\rm bli}$, Eq. \ref{eq:Sblinew}),
showing where the blister HII region is expanding when the association dies, where the
blister stalls, and where it produces a cometary cloud. For cometary clouds,
we assume that photoevaporation ceases at $t_{\rm com,max}={\rm min}(\tion, 2t_{\rm com})$.  Equation (\ref{eq:tion}) gives $\tion$, Equation (\ref{eq:t}) with $\xi_s=\xircom$ gives $\tcom$, and Equation (\ref{eq:t}) with
$\xi_s$ given by Equation (\ref{eq:xieq}) gives $t_{\rm stall}$. 
}
\label{Fig-E} 
\end{figure*}

\subsection{The Procedure for Analytic Solution for $M_{\rm loss}(t)$}

$M_{\rm loss}(t)$
is calculated with equations
\ref{eq:miios} to \ref{eq:mlosscom}, whose parameters are defined in
Table 1. $M_{\rm loss}(t)$ depends on the fixed parameters $\xi_{c0}$, $M_6$, $\Sigma_2$, and $S_{49}$.  The time dependence of $M_{\rm loss}$ comes from the
dependence of these equations  on
$\xircs(t)$ (Eq. \ref{eq:xis}), $\mu_{\rm r,cs}$ or $v_s(t)$ (Eq. \ref{eq:vs}).  The key timescales-- $t_{\rm stall}=t_s(\xi_{\rm stall})$, $\tcom=t_s(\xircom)$  and 
$\tion$--depend on the fixed input parameters. As shown in the basic
Equation (\ref{eq:mloss}), the mass loss, $M_{\rm loss}(t)$, can be broken into the sum of three individual mass loss terms, but the most complicated is $M_{\rm ion}(>\thecs)$.  The two simpler terms $M_{\rm init}(t)$ and $M_{\rm ion,os}(t)$
are given in Equations (\ref{eq:mej1}) and (\ref{eq:miios}) respectively.
The analytic solution follows these steps  to determine the three terms and find $M_{\rm loss}(t)$: 

1. Expanding blister  ($\tion< t_{\rm stall}$, $\tcom$). Use Equations (\ref{eq:xis}) and (\ref{eq:vs}) to follow $\xircs(t)$ and $v_s(t)$ from $\xircz$ for all $t$ up to $\tion$. Use Equation (\ref{eq:miong}) for $M_{\rm ion}(>\thecs)$. This analytic solution is an expanding blister solution for entire evolution (see upper right area of Fig. \ref{Fig-E}).

2. Expanding blister $\rightarrow$ Stalled blister ($t_{\rm stall}<\tion,\tcom$).  Use  step 1  for $0<t<t_{\rm stall}$. For $\tion>t>t_{\rm stall}$,  fix $\xi_{\rm cs,f}=\xi_{\rm stall}$ and use Equation (\ref{eq:stall}) to determine $M_{\rm ion}(>\thecs)$. This evolution is from expanding blister to stalled blister (left area of Fig. \ref{Fig-E}).

3.  Expanding blister $\rightarrow$ Cometary cloud ($\tcom<\tion,t_{\rm stall}$).   Use step 1 for $0<t<\tcom$ and use Equation (\ref{eq:mioncom}) for $\Delta M_{\rm com}(t)$  and Equation (\ref{eq:mlosscom}) for $M_{\rm loss}(t)$ for $t$ going from $\tcom$  up to $\tcomf={\min}( \tion, 2\tcom)$.  This evolution is from expanding blister to cometary cloud (bottom area of Fig. \ref{Fig-E}).

\newpage

\newpage

\eject

\label{lastpage}
\end{document}